\documentclass[manuscript]{aastex}

\usepackage{graphics,color}
\usepackage{epstopdf}
\definecolor{red}{rgb}{1,0,0}

\shorttitle{Asteroid Rotation Periods From iPTF}
\shortauthors{Chang et al.}
\begin{document}
\title{Asteroid Spin-Rate Study using the Intermediate Palomar Transient Factory}

\author{Chan-Kao Chang\altaffilmark{1}, Wing-Huen Ip\altaffilmark{1,2}, Hsing-Wen Lin\altaffilmark{1},
Yu-Chi Cheng\altaffilmark{1}, Chow-Choong Ngeow\altaffilmark{1}, Ting-Chang Yang\altaffilmark{1}, Adam
Waszczak\altaffilmark{3}, Shrinivas R. Kulkarni\altaffilmark{4}, David Levitan\altaffilmark{4},
Branimir Sesar\altaffilmark{4}, Russ Laher\altaffilmark{5}, Jason Surace\altaffilmark{5}, Thomas. A.
Prince\altaffilmark{4}}

\altaffiltext{1}{Institute of Astronomy, National Central University, Jhongli, Taiwan}
\altaffiltext{2}{Space Science Institute, Macau University of Science and Technology, Macau}
\altaffiltext{3}{Division of Geological and Planetary Sciences, California Institute of Technology, Pasadena, CA 91125, USA}
\altaffiltext{4}{Division of Physics, Mathematics and Astronomy, California Institute of Technology, Pasadena, CA 91125, USA}
\altaffiltext{5}{Spitzer Science Center, California Institute of Technology, M/S 314-6, Pasadena, CA 91125, USA}

\email{rex@astro.ncu.edu.tw}


\begin{abstract}
Two dedicated asteroid rotation-period surveys have been carried out using data taken on January 6-9
and February 20-23 of 2014 by the Intermediate Palomar Transient Factory (iPTF) in the $R$~band with
$\sim 20$-min cadence. The total survey area covered 174~deg$^2$ in the ecliptic plane. Reliable
rotation periods for 1,438 asteroids are obtained from a larger data set of 6,551 mostly main-belt
asteroids, each with $\geq 10$~detections. Analysis of 1751, PTF based, reliable rotation periods
clearly shows the ``spin barrier'' at $\sim 2$~hours for ``rubble-pile'' asteroids. We also found a
new large-sized super-fast rotator, 2005 UW163 \citep{Chang2014b}, and other five candidates as well.
Our spin-rate distributions of asteroids with $3 < D < 15$~km shows number decrease when frequency
greater than 5 rev/day, which is consistent to that of the Asteroid Light Curve Database
\citep[LCDB;][]{Warner2009} and the result of \citet{Masiero2009}. We found the discrepancy in the
spin-rate distribution between our result and \citet[][update 2014-04-20]{Pravec2008} is mainly from
asteroids with $\Delta m < 0.2$ mag that might be primarily due to different survey strategies. For
asteroids with $D \leq 3$~km, we found a significant number drop at $f = 6$ rev/day. The YORP effect
timescale for small-sized asteroid is shorter that makes more elongate objets spun up to reach their
spin-rate limit and results in break-up. The K-S test suggests a possible difference in the spin-rate
distributions of C- and S-type asteroids. We also find that C-type asteroids have a smaller spin-rate
limit than the S-type, which agrees with the general sense that the C-type has lower bulk density than
the S-type.
\end{abstract}

\keywords{surveys - minor planets, asteroids: general}

\section{Introduction}
Thanks to modern technology--wide-field detectors, high computing power, massive data storage, and
robotic observation, it is possible to obtain plenty of asteroid lightcurves within a short period of
time. Therefore, several important physical properties derived from asteroid lightcurves can be
investigated in a more comprehensive way. For instance, the asteroid rotation period ($P$) is the most
direct and essential property for understanding asteroids and the system they populate.

One of the key discoveries from asteroid rotation-period studies is the ``spin barrier'' at 2.2~hours
for asteroids of diameters ($D$) $>1$~km \citep{Harris1996}, which is crucial evidence for supporting
the ``rubble-pile'' model by which asteroids are believed to be made by gravitationally bounded
aggregations and would breakup under such super-fast rotation with $P < 2.2$~hr. \citet{Pravec2000}
take a step further to show that the spin barrier can hold for large-sized asteroids (i.e., $D >
150$~m), and other small-sized asteroids, as coherent monoliths, are able to rotate faster than that
limit \citep[i.e., super-fast-rotator; hereafter, SFR; see examples therein][]{Hergenrother2011}.

In fact, two large-sized SFRs have been discovered: 2001 OE84 \citep[$D \sim 0.65$~km, $P =
29.19$~min;][]{Pravec2002} and 2005 UW163 \citep[$D \sim 0.6$~km, $P = 1.29$~hr;][]{Chang2014b}. A
possible explanation for these large-sized SFRs is a size-dependent strength model for asteroids, in
which the cohesiveness and tensile strength of the asteroids decrease as its size increases.
Therefore, a smooth transition can occur between small monolithic asteroids and large rubble-pile
asteroids \citep{Holsapple2007}. In this way, a certain fraction of large-sized SFRs should be found
in asteroid populations. Identification of these large-sized SFRs along with their physical
properties, will provide definite answers about this special asteroid group.

Another notable discovery from analyzing asteroid rotation periods is that the spin-rate distribution
of smaller asteroids (i.e., $3 < D < 15$~km) have excesses both in the slow and fast ends. When an
asteroid system reaches collisional equilibrium, it exhibits a Maxwellian spin-rate distribution
\citep{Salo1987}, which has been seen for asteroids of $D > 40$~km \citep{Pravec2002}. Therefore, the
excesses of slow- and fast-rotating small asteroids become clear evidence of the
Yarkovsky-O'Keefe-Radzievskii-Paddack effect altering the spin rate of small-sized asteroids on
million year timescales \citep[YORP;][]{Rubincam2000}. However, the spin-rate distribution of
small-sized asteroids is still under debate. Instead of the flat distribution reported by
\citet{Pravec2002}, a non-flat distribution (i.e., more closely resembling a Maxwellian) was found by
\citet{Masiero2009}, \citet{Polishook2012} and \citet{Chang2014a}. Although the YORP effect can
account for both cases, a detailed investigation is still needed to understand the spin-rate evolution
of small asteroids.

Moreover, rubble-pile asteroids with a lower bulk density ($\rho$) should have a smaller spin-rate
limit \citep[$P \sim 3.3 \sqrt{(1 + \Delta m)/\rho}$; ][]{Pravec2000}. Consequently, the overall spin
rate of C-type asteroids (i.e., $\rho \sim 1.33 \pm 0.58$~g/cm$^3$) should be lower than for the
S-type (i.e., $\rho \sim 2.72 \pm 0.54$~g/cm$^3$) \citep{Demeo2013}. Although \citet{Chang2014a}
attempted to investigated whether this trend exists between various spectral-type asteroids, the
conclusion was still preliminary due to only a few tens of objects available in their samples.

To gain a more comprehensive understanding with respect to the aforementioned questions, we continued
our previous study and conducted two more asteroid rotation-period surveys of data taken on January
6-9 and February 20-23 of 2014 by the iPTF\footnote{Intermediate Palomar Transient Factory;
http://ptf.caltech.edu/iptf}.  These surveys included more samples and a higher level of completeness.
The total survey area was $\sim 174$~deg$^2$ over the ecliptic plane, and 6,551 asteroids of $\geq
10$~detections were obtained. From these lightcurves, 1,438 reliable rotation periods were derived and
among them we found six large-sized SFR candidates, one of which, (335433) 2005 UW163, was confirmed
via follow-up observation \citep{Chang2014b}.

We describe the observation information and lightcurve extraction in Section 2. The periodicity
analysis is given in Section 3. The results and discussion are presented in Section 4. A summary and
conclusion can be found in Section 5.

\section{Observations and Data Reduction}
The PTF project, and its successor iPTF project, employs the Palomar 48-inch Oschin Schmidt Telescope
equipped with an 11-chip mosaic CCD camera to explore the transient and variable sky synoptically.
This configuration creates a field of view of $\sim7.26$ \,deg$^2$ and a pixel scale of
1.01\arcsec~\citep{Law2009, Rau2009}. Most PTF exposures are taken in the Mould-{\it R} band, and the
other available filters are the Gunn-{\it g\arcmin} and two different $H_{\alpha}$ bands.  A 60-second
$R$-band exposure can reach a median limiting magnitude of $\sim$21 mag at the $5\sigma$ level
\citep{Law2010}.

Each PTF exposure is processed by the IPAC-PTF photometric pipeline, which includes image splitting,
de-biasing, flat-fielding, source extraction, astrometric calibration, and photometric calibration to
generate reduced images and source catalogs \citep{Grillmair2010, Laher2014}. The absolute magnitude
calibration is done against Sloan Digital Sky Survey fields \citep[hereafter, SDSS;][]{York2000} on a
per-night, per-filter, per-chip basis, and routinely reaches a precision of $\sim0.02$~mag on
photometric nights \citep{Ofek2012a, Ofek2012b}. Small photometric zero-point variations are possible
between catalogs of different nights, fields, filters and chips. Its accuracy depends on the degree
affected by weather and transient variations in atmospheric conditions within a night.

\begin{figure}
  \plotone{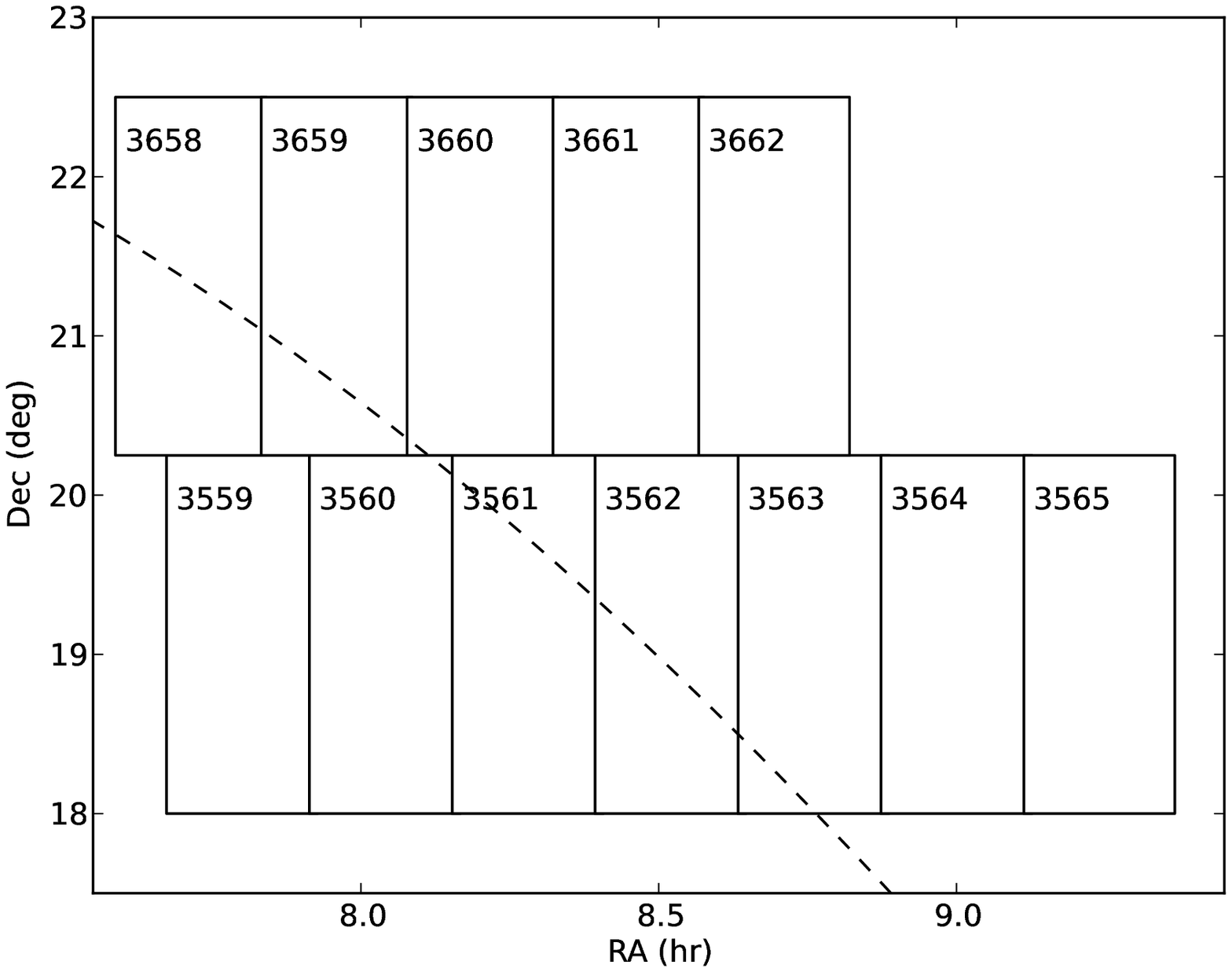}
  \caption{Field configurations for survey data taken in January 2014.
  The rectangles show PTF fields with corresponding field ID. The dashed line shows the position of the ecliptic  plane.
  Note that the scales of right ascension and
  declination are different.}
  \label{obs_fig}
  \end{figure}

\begin{figure}
  \plotone{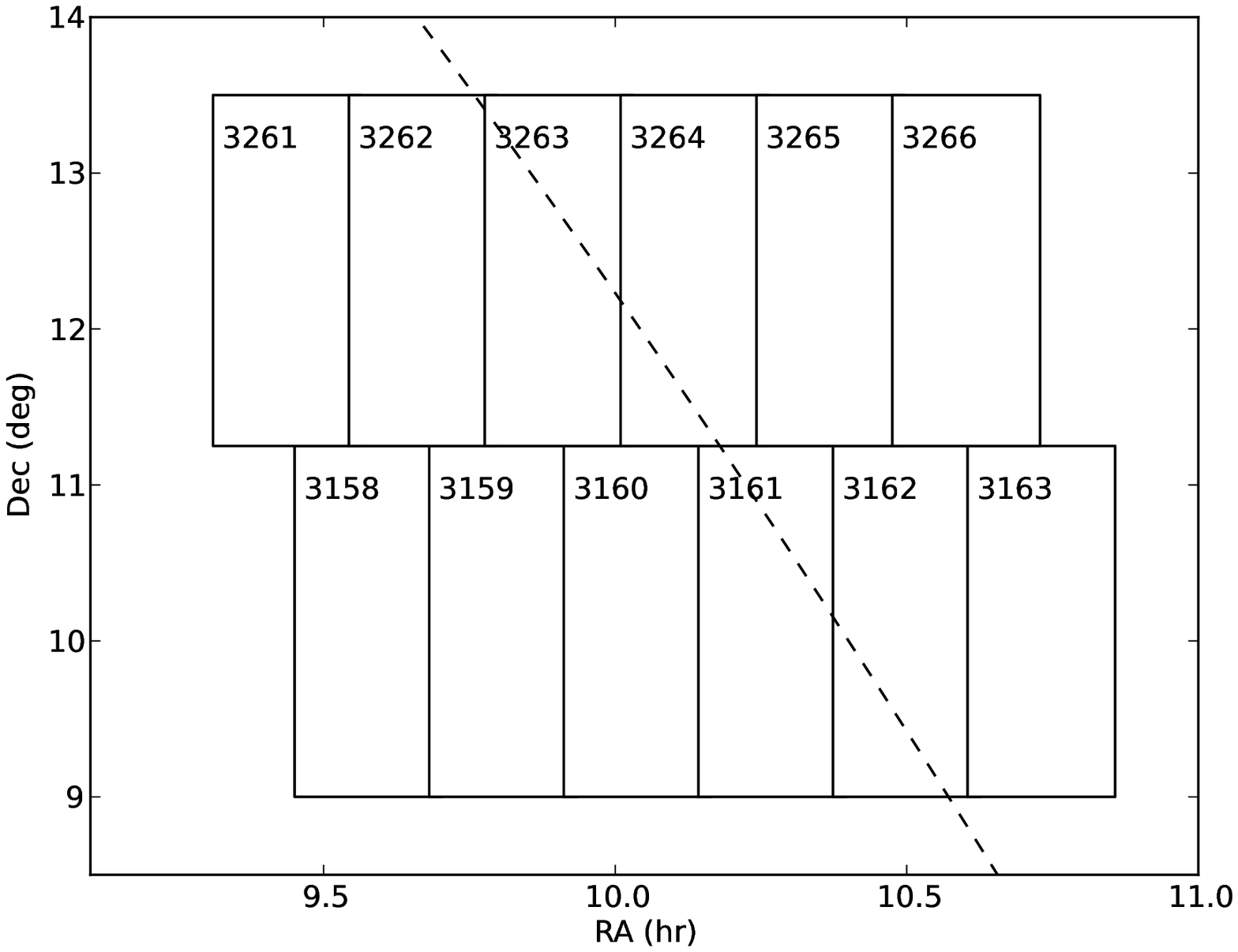}
  \caption{Field configurations for survey data taken in February 2014.
  The rectangles show PTF fields with corresponding field ID. The dashed line shows the position of the ecliptic  plane.
  Note that the scales of right ascension and
  declination are different.}
  \label{obs_fig1}
  \end{figure}

To collect a large sample size of asteroid lightcurves for furthering our previous asteroid spin-rate
studies \citep{Polishook2012, Chang2014a}, we conducted two asteroid rotation-period surveys during
January 6-9 and February 20-23 in 2014. Each survey continuously scanned twelve consecutively numbered
PTF fields in the ecliptic plane in the $R$-band with a cadence of $\sim 20$~min. The exposure time of
each frame was 60~s, and the scanned sky coverage was $\sim 174$~deg$^2$ in total. The observational
metadata are given in Tables~\ref{obs_log_01}, and the field configurations are shown
Fig.~\ref{obs_fig} and ~\ref{obs_fig1}.  After purging all stationary sources, the source catalogs
were matched against the ephemerides obtained from the {\it JPL/HORIZONS} system with a radius of
2\arcsec~to extract the lightcurves of known asteroids.  We also excluded any detection flagged as a
defect by the IPAC-PTF photometric pipeline from the lightcurves. In the end, we were left with 6,551
asteroid lightcurves, each with $\geq 10$~data points (hereafter, PTF-detected asteroids) for the
following rotation-period analysis.

\section{Rotation-Period Analysis}\label{period_analysis}
Before measuring rotation period, the orbital elements obtained from the Minor Planet
Center\footnote{http://minorplanetcenter.net} were used to correct for light-travel time and reduce
magnitudes to both heliocentric, $r$, and geocentric, $\triangle$, distances at 1~AU for all
PTF-detected asteroids. Moreover, the absolute magnitudes, $H_R$, were simply estimated by applying a
fixed $G_R$ slope of 0.15 in the $H$--$G$ system \citep{Bowell1989} due to the small change in phase
angles within our four-night observations.

We followed the traditional second-order Fourier series method to fit our lightcurves
\citep{Harris1989}:
\begin{equation}\label{FTeq}
  M_{i,j} = \sum_{k=1,2}^{N_k} B_k\sin\left[\frac{2\pi k}{P} (t_j-t_0)\right] + C_k\cos\left[\frac{2\pi k}{P} (t_j-t_0)\right] + Z_i,
\end{equation}
where $M_{i,j}$ is the $R$-band reduced magnitude measured at the light-travel time corrected epoch
$t_j$, $B_k$ and $C_k$ are the Fourier coefficients, $P$ is the rotation period, and $t_0$ is an
arbitrary epoch. Here, we also introduced a constant value $Z_i$ to correct the small photometric
zero-point variations between data acquired from different nights, fields, filters and chips. Then,
Eq.~(\ref{FTeq}) was solved by using least-squares minimization for each given $P$ to obtain other
free parameters. The spin-rate ($f$) range was stepped through from 0.25 to 25~rev/day (e.g., about 1
to 96~hr) with increments of 0.025~rev/day.

The results were reviewed and a quality code ($U$) was manually assigned to each folded lightcurve,
where: `3'~means highly reliable; `2'~means some ambiguity; `1'~means possible, but may be wrong
\citep{Warner2009}. Moreover, when the lightcurve was unable to find any acceptable solution, it was
assigned $U = 0$. The uncertainty of the derived rotation period was estimated from periods having
$\chi^2$ smaller than $\chi_{best}^2+\triangle\chi^2$, where $\chi_{best}^2$ is the $\chi^2$ of the
derived period and $\triangle\chi^2$ is calculated from the inverse $\chi^2$ distribution, assuming $1
+ 2N_k + N_i$~degrees of freedom. The amplitude was calculated from the peak-to-peak variations after
rejecting the upper and lower 5\% of data points to avoid outliers, which are probably contaminated by
nearby bright stars or unfiltered artifacts from the lightcurve extraction.

\section{Results and Discussion}
Most PTF-detected asteroids are main-belt asteroids, and the others include Hilda, Jovian Trojan, and
near-Earth objects. If the object are available in $WISE$/$NEOWISE$ data set \citep{Grav2011,
Mainzer2011, Masiero2011}, then we adopted their diameter estimation. Otherwise, the diameter was
estimated using
\begin{equation}\label{dia_eq}
  D = {1130 \over \sqrt{p_R}} 10^{-H_R/5},
\end{equation}
where $D$ is the diameter in~km, $p_R$ is the $R$-band geometric albedo, and 1130 is the conversion
constant adopted from \citet{Jweitt2013}. We assumed three empirical albedo values, $p_R =0.20$, 0.08
and 0.04, for objects in inner ($2.1 < a < 2.5 AU$), mid ($2.5 < a < 2.8$ AU) and outer ($a > 2.8$ AU)
main belts, respectively \citep{Tedesco2005}.

\subsection{Derived Rotation Periods}\label{discuss_p}
\begin{figure}
  \plotone{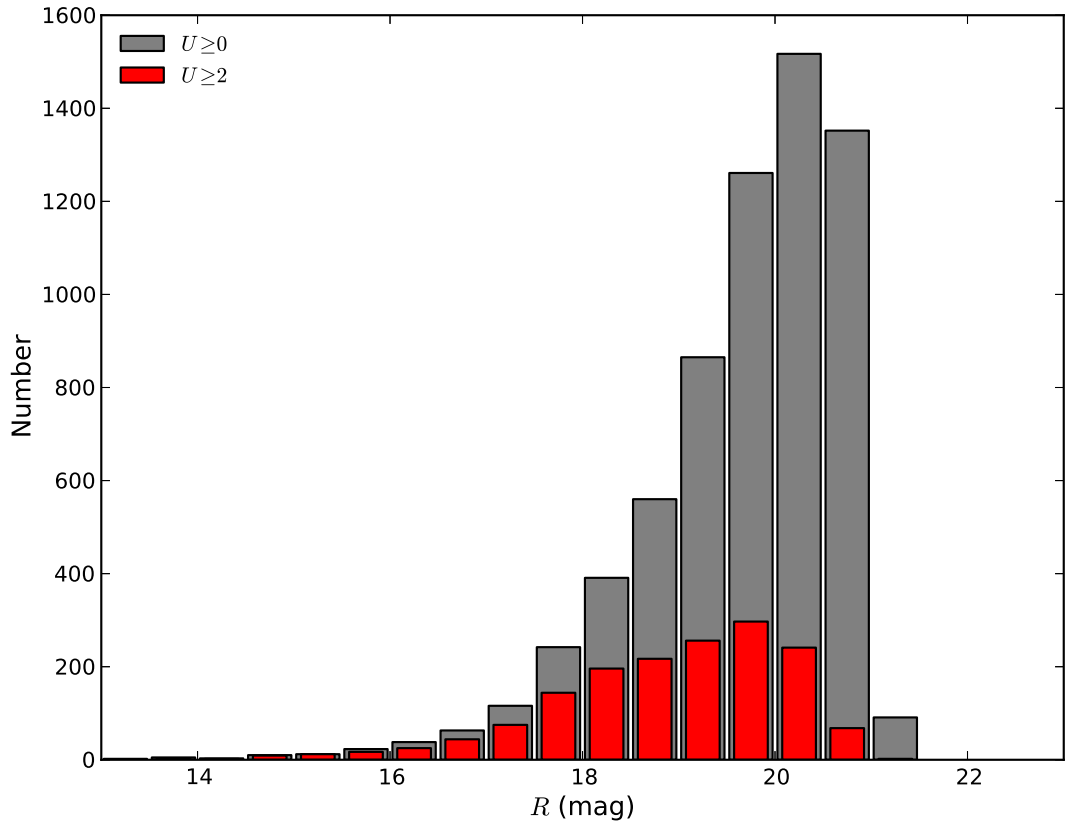}
  \caption{Magnitude distributions of PTF-detected asteroids (gray, $U >0$) vs.\ PTF-U2s (red, $U \geq 2$).
  (Values of quality code $U$ are defined in Section~\ref{period_analysis}.) }
  \label{mag_his}
  \end{figure}

\begin{figure}
  \plotone{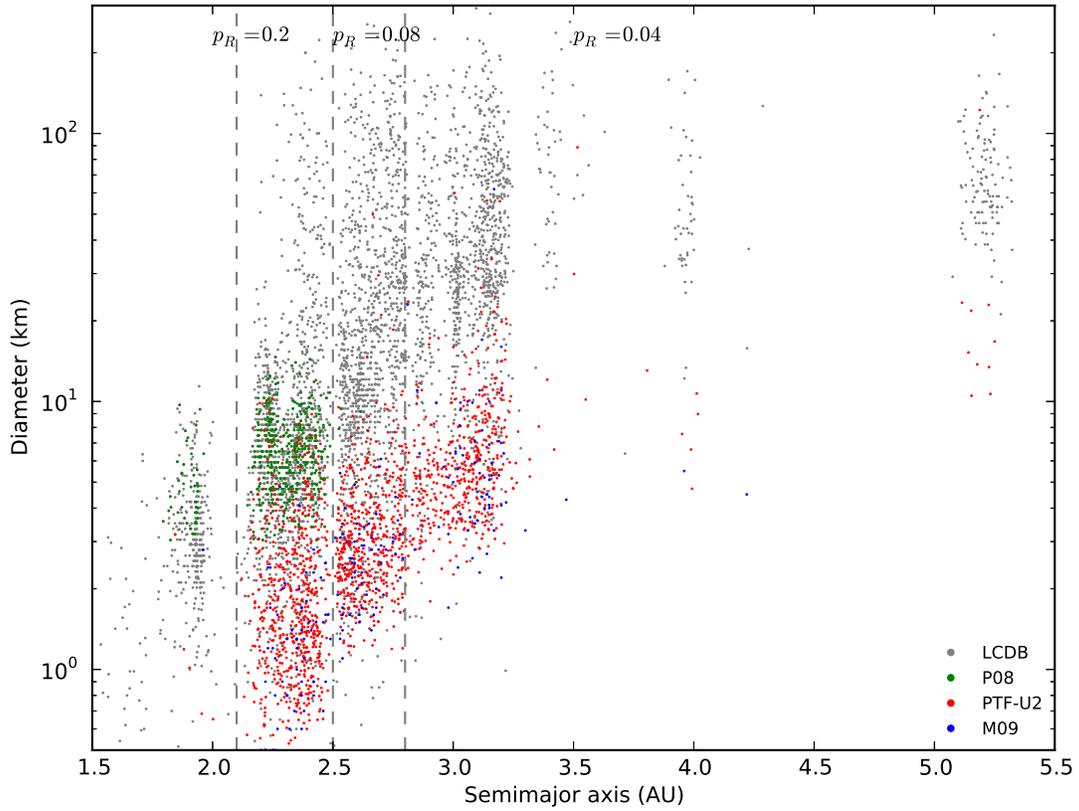}
  \caption{Diameter vs.\ semi-major axes for PTF-U2s (red), P08 (update 2014-04-20, green), M09 (blue) and LCDB (gray).
  The dashed lines show the divisions of empirical geometric albedo ($p_R$) for asteroids located at
  different regions of the semi-major axis.}
  \label{a_d}
  \end{figure}

\begin{figure}
\includegraphics[angle=90,scale=.7]{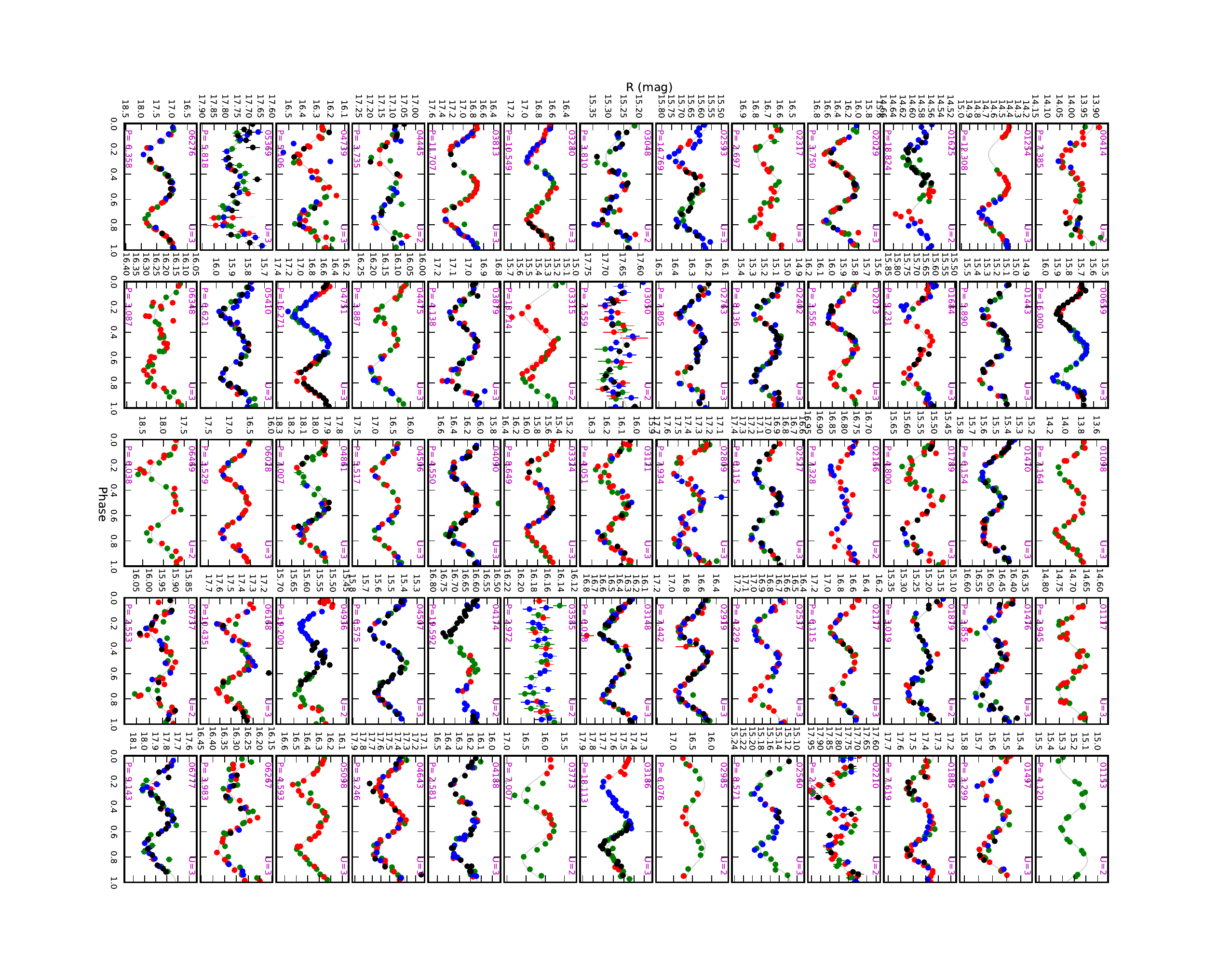}
\caption{Set of 65 folded lightcurves for the PTF-U2 asteroids. The green, red, blue and black
circles represent observation taken at different nights. The asteroid designation is given
on each plot along with its derived rotation period $P$ in hours and quality code $U$.}
\label{lightcurve00}
\end{figure}

\begin{figure}
\includegraphics[angle=90,scale=.7]{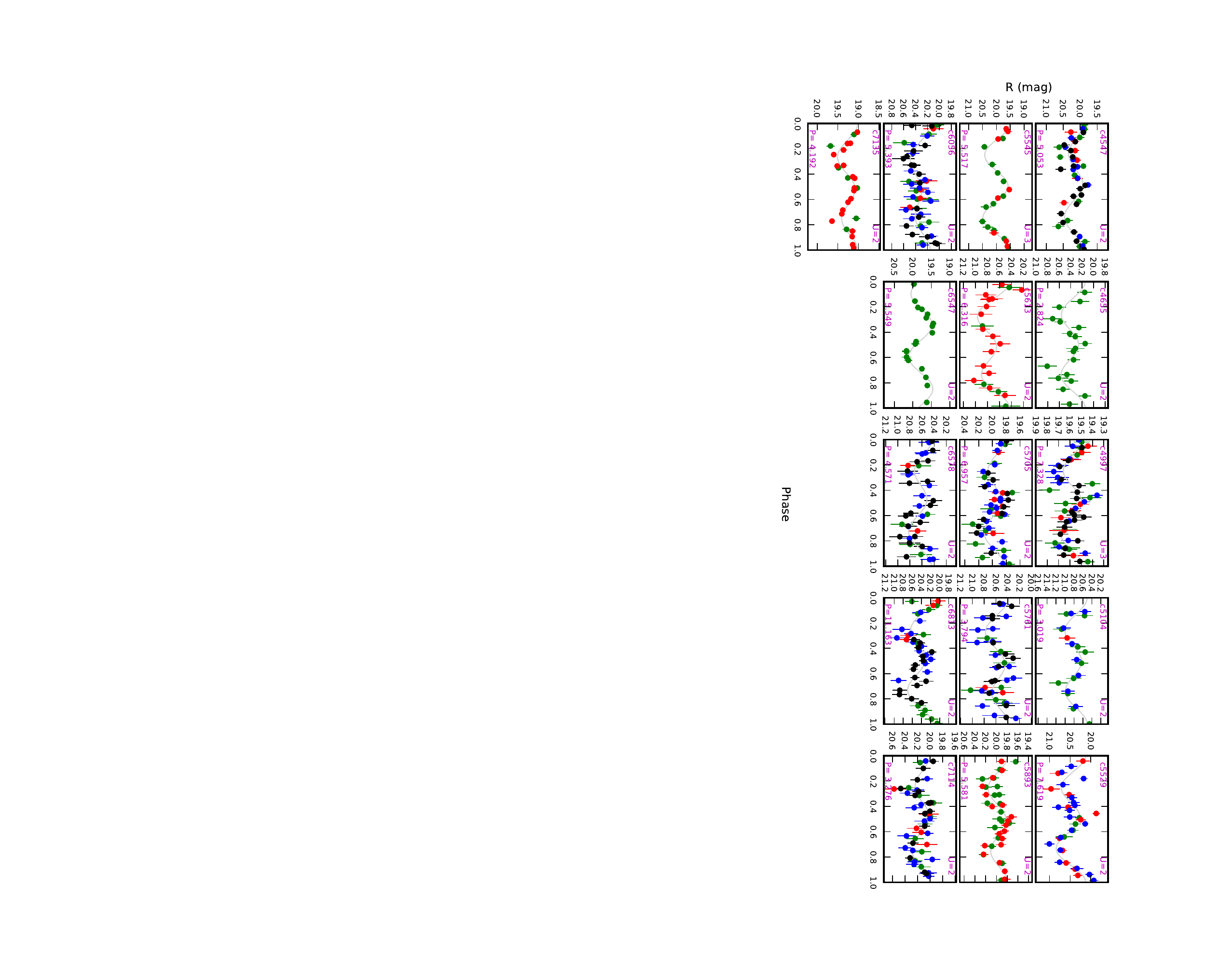}
\caption{Same as Fig.~\ref{lightcurve00} for other 15 PTF-U2 asteroids.}
\label{lightcurve21}
\end{figure}

From PTF-detected asteroids, we obtained 1,438 reliable (i.e., $U \ge 2$) rotation periods (hereafter,
PTF-U2s). The magnitude distributions of the PTF-detected asteroids vs. PTF-U2s are shown in
Fig.~\ref{mag_his}, which peak at $\sim 20.5$ and $\sim 19.5$~mag, respectively. Fig.~\ref{a_d} shows
the plot of the semi-major axis vs. diameter for PTF-U2s, where we see the low-limit diameters
increasing as a function of semi-major axis. The derived rotation periods and lightcurve amplitudes,
along with the orbital elements and observational conditions, are summarized in Table~\ref{table_p},
and their folded lightcurves are provided in Figs.~\ref{lightcurve00}-\ref{lightcurve21}.

\begin{figure}
\includegraphics[angle=90,scale=.7]{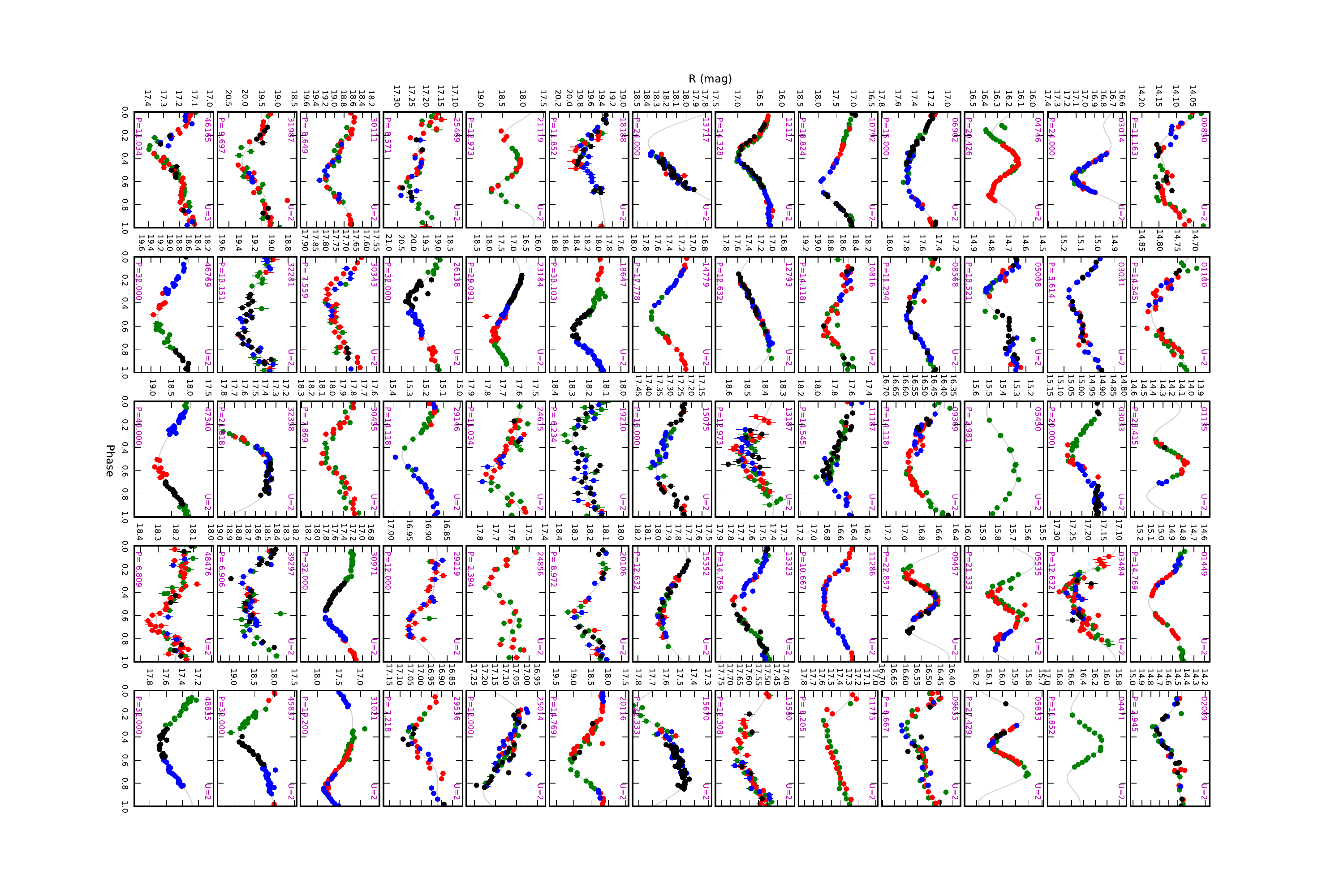}
\caption{Same as Fig.~\ref{lightcurve00} for other 65 PTF-P asteroids.}
\label{lightcurve_p_00}
\end{figure}

In addition, there are 169 objects having folded lightcurves with partial coverage in rotational
phase, but nevertheless a clear trend (hereafter, PTF-Ps). Most PTF-Ps seem to be long-period objects
(i.e., $f < 2$ rev/day), and therefore we only can obtain their partial lightcurves. We estimated a
lower limit on the amplitude based on the lightcurve variation and their actual rotation period should
be longer than the derived. These objects are summarized in Table~\ref{table_p_part} and their folded
lightcurves are given in Fig.~\ref{lightcurve_p_00}-\ref{lightcurve_p_02}.

\begin{figure}
\includegraphics[angle=90,scale=.7]{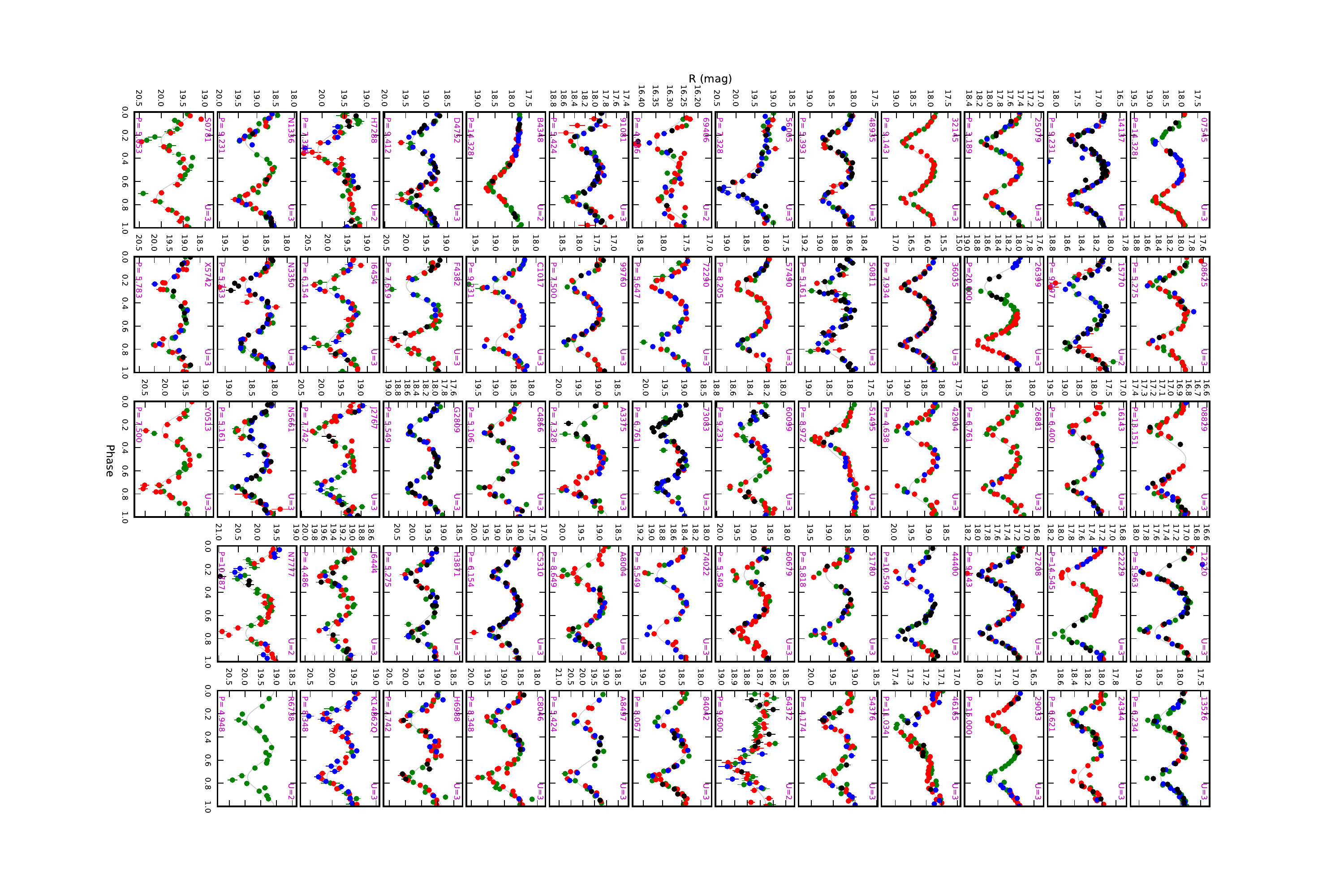}
\caption{Same as Fig.~\ref{lightcurve00} for 63 asteroids with large-amplitude lightcurves and a deep V-shape minima.}
\label{lightcurve_b_00}
\end{figure}

We also found 63 asteroids showing large amplitudes and deep V-shaped minima in their lightcurves (see
Fig.~\ref{lightcurve_b_00}). Most of them are probably contact binaries or very elongate asteroids,
which have smooth transitions from minimum to maximum \citep[see 3169 Ostro as an example
in][]{Descamps2007}. Among them, we use the typical features of lightcurve for binary asteroid, a deep
V-shaped minimum with a ``shoulder''  due to the abrupt drop in brightness at the onset of
eclipse/occultation and an inverse U-shaped maxima \citep{Behrend2006}, to identify several binary
asteroid candidates, which are 51495 and 56005, and maybe 46165 and B4348. These candidates need to be
confirmed by follow-up observations. Moreover, the asynchronous binary candidate, (69406) 1995 SX48
\citep{Warner2014}, was also observed in our survey, and we detected a primary rotation period of
$\sim$4.49 hr from our relatively scattered lightcurve.

\begin{figure}
  \plotone{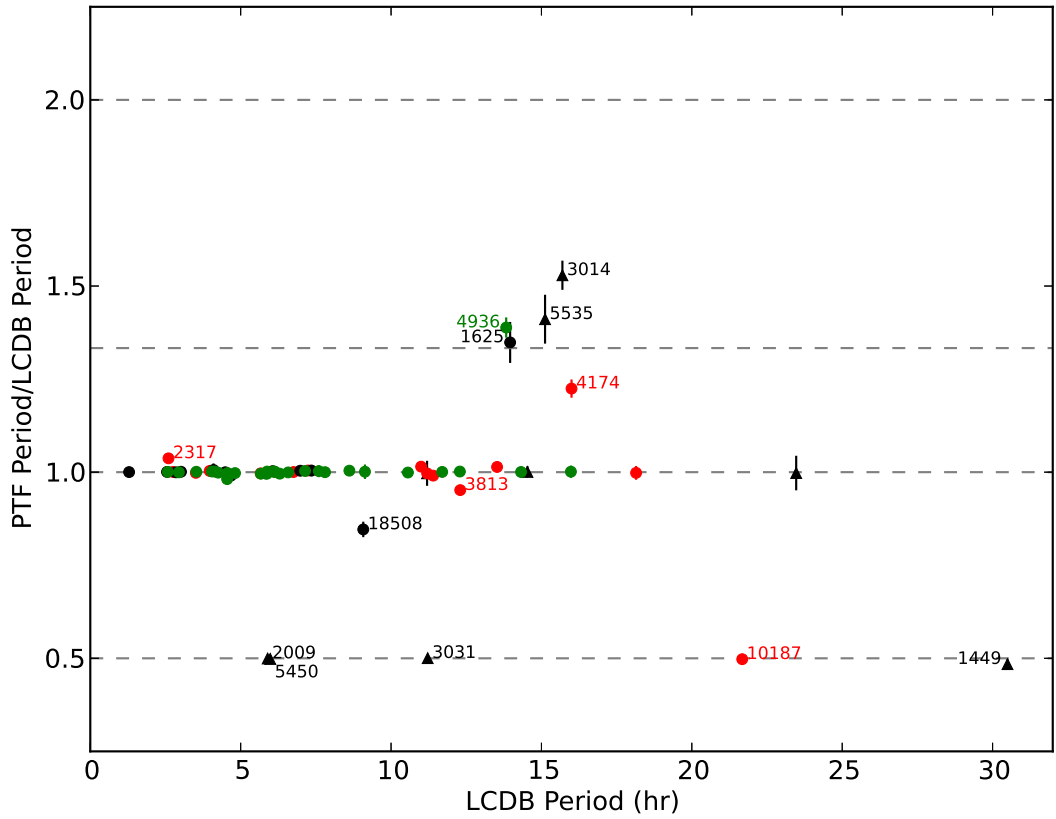}
  \caption{Comparison of 53 derived rotation periods for PTF-U2s with
    matching objects in the Asteroid Light Curve Database.  Filled
   circles and filled triangles correspond to
   asteroids with full and partial rotational-phase coverages,
   respectively.  Red, green, and black indicate
  $U$ of the PTF-U2 is $>$,  $=$, or $< U$ of the matching LCDB object, respectively.}
  \label{diff_p}
  \end{figure}

Among the 1,438 PTF-U2s, 65 objects have published rotation periods in the Asteroid Light Curve
Database \citep[LCDB;][]{Warner2009}\footnote{http://www.minorplanet.info/lightcurvedatabase.html}. To
ensure the reliability of our period analysis, we compared the rotation periods between our survey
data and those of matching objects in the LCDB, and show the results in Fig.~\ref{diff_p}. Most of the
matches have consistent derived rotation periods. However, there are 13 outliers. Despite the six
PTF-Ps (1449, 2009, 3014, 3031, 5450 and 5535) that are relatively uncertain in our determination of
rotation periods and the four (2317, 3813, 4174 and 10187) that have better quality codes in our
results than in the LCDB, there are still three outliers (1625, 4936 and 185086) having rotation
periods that are significantly different from the rotation period of its LCDB counterpart within a
reasonable uncertainty. The possible reasons are discussed below.

Asteroid 1625 has a long rotation period of 13.96~hr with $U = 3$ \citep{Higgins2011}. We were only
able to obtain part of its full lightcurve, and thus derived a rotation period of 18.82~hr, which is
$\sim 4/3$ of the corresponding LCDB rotation period (see Fig.~\ref{lightcurve01}).

Asteroid 4936 has $U = 2$ ratings from both our determination and the LCDB. When the PTF lightcurve is
folded with the rotation period of 13.83~hr listed in the LCDB, it does not give a clear trend.
However, folding the PTF lightcurve with a rotation period of 19.2~hr gives a much better result, one
that looks better, in fact, than the similarly folded lightcurve for this asteroid reported by
\citet{Pray2008}.  Therefore, we believe our rotation period is more accurate than the LCDB rotation
period.

Asteroid 185086 also has a long rotation period of 9.08~hr with $U = 3$ \citep{Masiero2009}.  It is
also a faint object (see Fig.~\ref{lightcurve15}), which is one possible reason for why we were only
able to obtain only a sparse lightcurve, and therefore make a less accurate determination of its
rotation period.

Among the 6,551 PTF-detected asteroids, 96 objects have $R < 19$~mag (i.e., large-sized) and $\Delta m
> 0.5$~mag, but do not have a rotation-period determination. These asteroids are listed in
Table~\ref{table_nd}. The associated lightcurves show a long-trend variation over our four-night
observing-time span, suggesting that they very likely have a spin rate of $<1$~rev/day. In order to
obtain their relatively long rotation periods, observations with a longer time baseline are required,
which is beyond the capability of our present survey strategy. Thus, we cannot more accurately
determine the spin rates of these large-sized, long-period asteroids.

\subsection{Detecting Simulation for Asteroid Rotation Period}
We adopted similar approach described in \citet{Masiero2009} to carry out the detecting simulation for
asteroid rotation period. The apparent magnitude ($m$) distribution of PTF-detected asteroids can be
described as
\begin{equation}
  N = {2.5^{(M-12)} \over 1+e^{(M-20.5)/0.25}}.
\end{equation}
This function accounts for the number increase and falloff along with apparent magnitude in a
magnitude-limited survey \citep{Jedicke1997}. Instead of using sophisticated asteroid shape to
construct synthetic lightcurve that introduces formidable time-consuming computation, the synthetic
objects were assumed relaxed triaxial ellipsoids having equal axes in $b$ and $c$ and rotating around
the principle-axis $a$. Therefore, the lightcurves can be written as
\begin{equation}
  m = 2.5 \log_{10} \sqrt{1 + \left [ \left ( b \over a \right )^2 -1 \right ] \cos^2(2\pi\phi)\sin^2\theta},
\end{equation}
and the amplitude ($\Delta m$) can be written as
\begin{equation}
  \Delta m = 2.5 \log_{10} \left [ \cos^2\theta + \left ( b \over a \right )^2 \sin^2\theta \right ]^{-1/2},
\end{equation}
where $\phi$ is rotational phase and $\theta$ is the angle between line of sight and spin vector
\citep{Lacerda2003, Lacerda2007}. The cadences of the synthetic lightcurves were chosen to be
identical with the survey observations and the numbers of synthetic measurements were assigned
according to the number of detections of PTF-detected asteroids with different apparent magnitudes.

We generated 402,000 synthetic lightcurves that uniformly distribute in frequency of $0 \le f \le 12$
rev/day, in pole orientation of $10 \le \theta \le 90$ degree and in axis ratio of $0.1 \le b/a \le
1$. Then, these synthetic lightcurves were analyzed by the aforementioned second-order Fourier series
method to search for assigned periods. We defined a successful period determination as : (a) the
derived period is within 5\% of the original period, (b) the folded lightcurve of the derived period
has double peaks or double dips and (c) the assigned photometric error is smaller than the derived
amplitude. Therefore, the ``PTF-P''-like objects could not be picked up in the simulation.

\begin{figure}
  \plotone{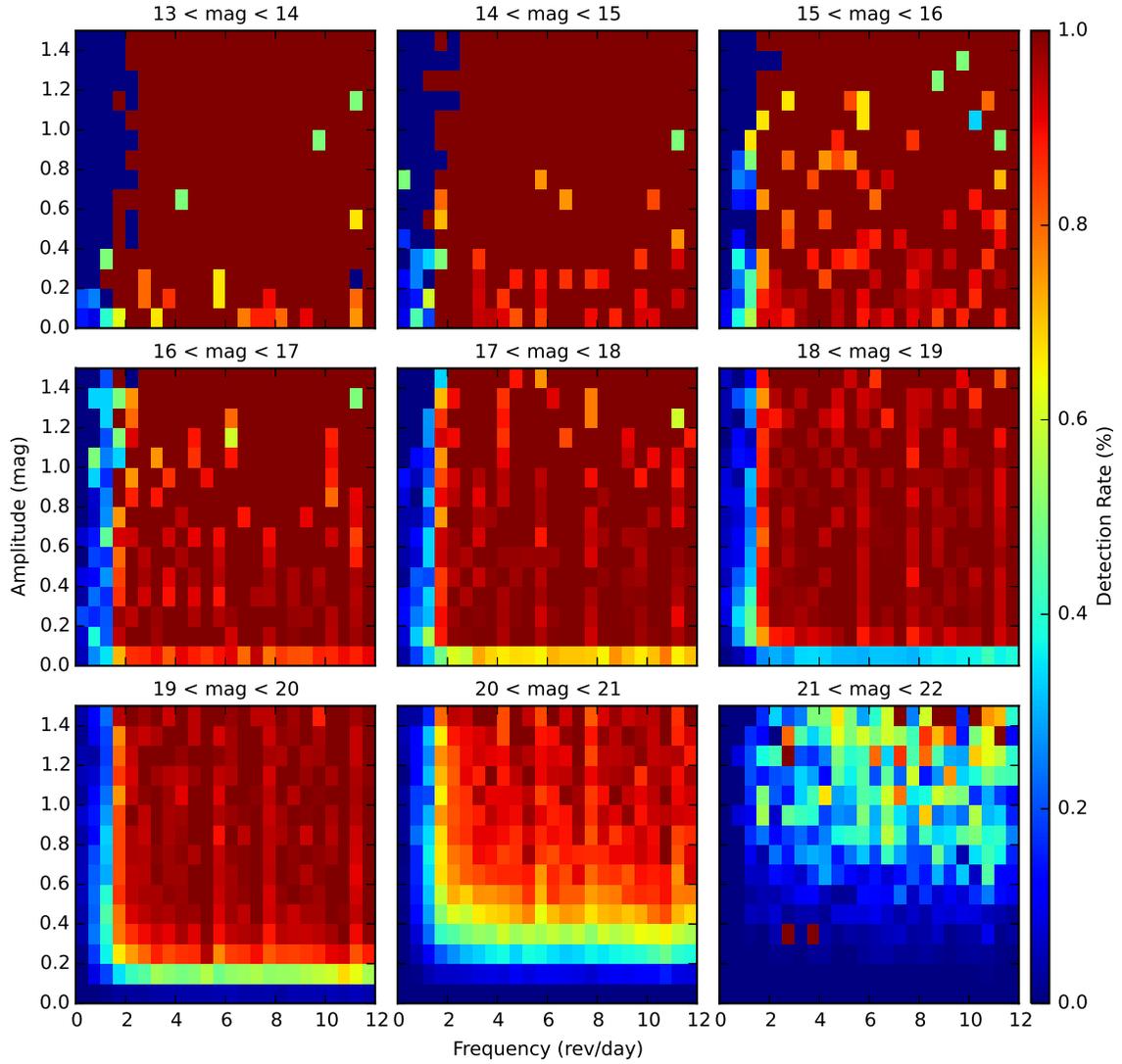}
  \caption{Detection rates for asteroid rotation period. The color bar scale on the right shows the percentage
  of successful recovery for rotation period of synthetic objects. The apparent magnitude interval is indicated
  on the top of each plot.}
  \label{debias_map}
  \end{figure}

Fig.~\ref{debias_map} shows the detection rate of $f$ vs. $\Delta m$ for various apparent magnitude
intervals. In general, the detecting efficiency increases with $\Delta m$ and decreases with $m$.
Moreover, the detection rate becomes very low for long-period objects that we see an obvious drop off
at $f < 1.5$ rev/day. When the assigned photometric error is comparable with $\Delta m$ for a
synthetic object, the simulation would not able to detect its rotation period. Therefore, the chance
of detecting small-amplitude objects becomes much lower for faint objects. However, we do not see any
favoring of detecting particular frequency with fixed lightcurve amplitude except for long-period
objects (i.e., $f < 1.5$ rev/day). Then, we applied these detection rate to debias our result with
interval of $\Delta m = 0.1$ mag, $f = 0.5$ rev/day and $m = 0.5$ mag. It is worth to note that the
triaxial ellipsoid used in the simulation could not fully represent the sophisticated asteroid shape
that would overestimate the detecting rate especially for low-amplitude object. However, including all
possible asteroid shapes in such simulation to investigate the biases from various factors in a
detailed way would becomes a formidable computing process that is far beyond the scope of this study.
Our simulation can be the first step to understand whether the survey and the analysis tend to detect
particular frequencies.

\subsection{Statistical Analysis}\label{statisticalanalysis}
\subsubsection{Spin-Rate Limit and Mean Spin Rate}
\begin{figure}
  \plotone{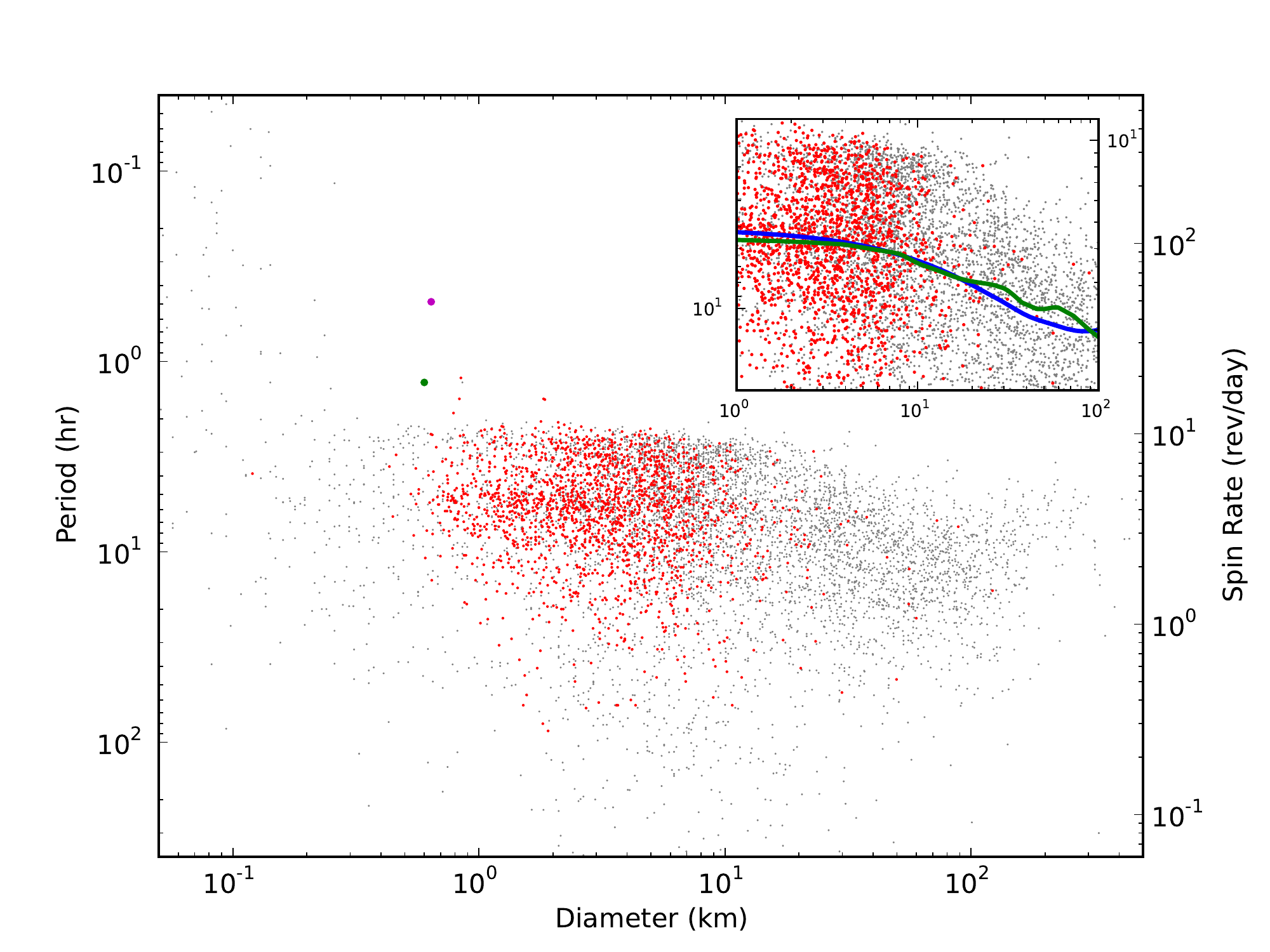}
  \caption{Asteroid rotation period vs.\ diameter. The red and gray filled circles are PTF-U2s and LCDB objects
  of $U \geq 2$, respectively. The SFRs, 2001 OE84 (magenta) and 2005 UW163 (green), are indicated with larger
  filled circles. The inset plot is a zoom-in of the dense region, where the green and blue lines
  are the regressions of the spin rates for PTF-U2s and LCDB objects computed using LOWESS, respectively.}
  \label{dia_per}
  \end{figure}

\begin{figure}
  \plotone{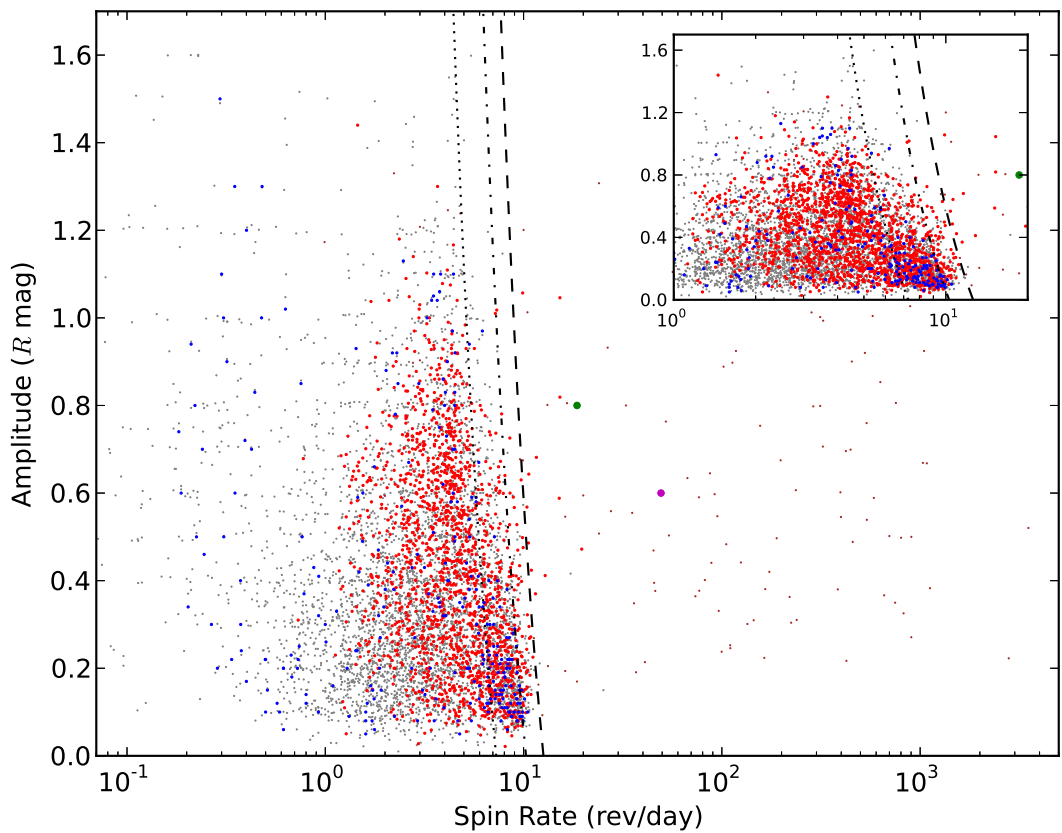}
  \caption{Lightcurve amplitude vs.\ spin rate. The red, gray, blue, and brown filled circles are PTF-U2s,
  LCDB of $D \geq 0.2$~km, P08 (update 2014-04-20), and LCDB of $D < 0.2$~km, respectively.
  The dashed, dot-dashed and dotted lines represent the spin-rate limits for rubble-pile asteroids
  with bulk densities of 3, 2, and 1~g/cm$^3$ \citep{Pravec2000},
  respectively.  The inset plot is a zoom-in of $1 < f < 12$~rev/day.}
  \label{spin_amp}
  \end{figure}

\begin{figure}
\includegraphics[angle=0,scale=.7]{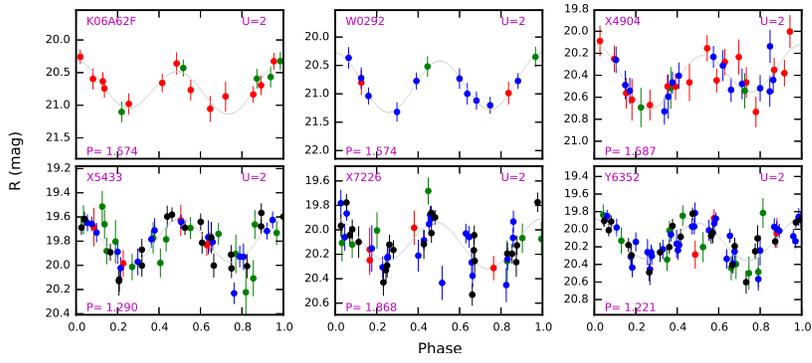}
\caption{Same as Fig.~\ref{lightcurve00} for 6 SFR candidates, in which X5433 has been
confirmed its super-fast rotating.}
\label{lightcurveSFR}
\end{figure}

To compare PTF results obtained both from this work and \citet{Chang2014a}, i.e., 1,751 PTF-U2s in
total, with the objects of $U = 2$ in the LCDB, Fig.~\ref{dia_per} shows the plot of diameter vs.\
rotation period for objects having quality code $U = 2$. Because of our four-night observing-time span
and limiting-magnitude range, most PTF-U2s are confined to the region of $2 < P < 20$~hr and $1 < D <
10$ km, where most LCDB objects populate. The 2.2~hr spin barrier is clearly seen for objects of $D >
150$~m, which indicates the spin-rate limit for rubble-pile asteroids under self-gravity. A small
group of monolithic SFRs located at $D < 150$~m and $P < 2.2$~hr, which can survive fast rotation
without breakup, due to other mechanical forces besides self-gravity. In addition to 2001 OE84, we
also found six large SFRs candidates, (320292) 2007 RO221, (334904) 2003 WL167, (335433) 2005 UW163,
(337226) 2000 EL98, (346352) 2008 RM118, and 2006 AF62 (see Fig.~\ref{lightcurveSFR}), in which the
super fast rotation of 2005 UW163 has been confirmed \citep{Chang2014b}. While we plot the spin rate
vs.\ amplitude in Fig.~\ref{spin_amp}, all SFRs have bulk density $\rho > 3$~g/cm$^3$ if $P \sim 3.3
\sqrt{(1+\Delta m)/\rho}$ is applied for rubble-pile asteroids, where $\Delta m$ is the lightcurve
amplitude \citep{Pravec2000}.  Such high bulk density is very unusual among asteroids. The
size-dependent strength model \citep{Holsapple2007} provides an alternative explanation, and we thus
should observe a certain fraction of large SFRs. Although the five unconfirmed SFR candidates show
reasonably good folded lightcurves, their super fast rotations still need to be verified by follow-up
observations to exclude the possibility of noise-induced false-positive detections \citep{Harris2012}.
If the SFR population is consistent with the overall asteroid spin-rate distribution, then we can rule
them out as a distinct asteroid group.

The inset plot of Fig.~\ref{dia_per} is a detailed view of the dense region, where the regressions of
the spin rate for PTF-U2s and LCDB are computed using locally weighted scatterplot smoothing
\citep[LOWESS;][]{Cleveland1979}. Both regressions share a similar trend that is flat for small-sized
asteroids and then gradually changes to longer rotation periods for larger-sized asteroids.

\subsubsection{Spin-Rate Distribution}
The spin-rate distribution for small-sized asteroids is important for understanding the evolution of
asteroid systems. At the moment the available catalogs with large data volume include
\citet[][]{Pravec2008}, \citet[][hereafter, M09]{Masiero2009} and the LCDB. \citet[][]{Pravec2008}
have been collecting asteroid spin rate for more than a decade; therefore, the version of April 20
2014 of their data set containing 462 asteroids is used in the following analysis (hereafter, P08,
update 2014-04-20; private communication). We also include PTF-Ps in the spin-rate distribution of
PTF-U2s by doubling their derived rotation period as suggested by their folded lightcurve. $\sim$ 95
\% of them are of $f \leq 2$ rev/day, and the contribution of the rest is is very limited with respect
to the whole PTF-U2s. In addition, the detecting simulation does not pick up the ``PTF-P''-like
objects that results in an underestimated detection rate for $f \le 2$ rev/day and consequently an
overestimation for long-period asteroids in the debiased result. Therefore, these two bins have
relatively large uncertainty that we see them as a reference and not considered in the following
discussion.

\begin{figure}
  \plotone{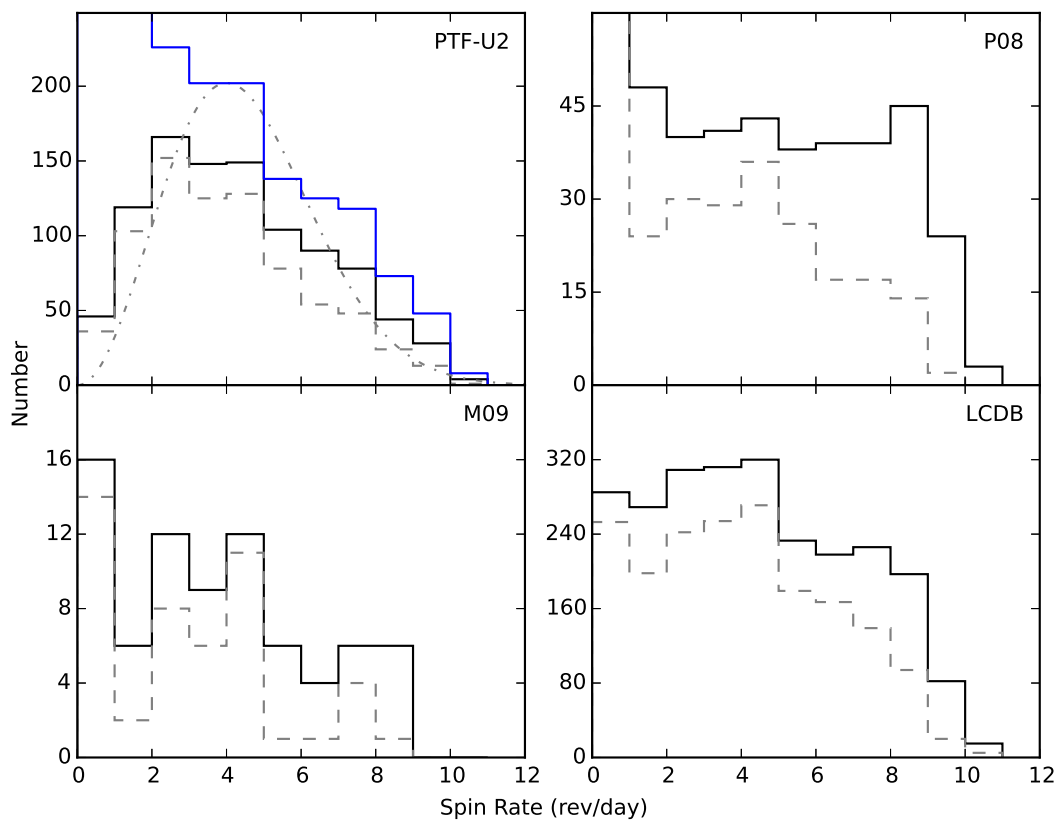}
  \caption{Spin-rate distributions of D3-15 asteroids for PTF-U2s, P08 (update 2014-04-20), M08 and LCDB.
  The gray dashed line are for asteroids of $\Delta m > 0.2$~mag, the gray dot-dashed line on the
  plot for PTF-U2s is the best-fit Maxwellian distribution and the blue line is debiased result for PTF-U2s.
  Note that P08 (update 2014-04-20) is not in full scale.}
  \label{spin_rate_comp}
  \end{figure}

\begin{figure}
  \epsscale{1}
  \plotone{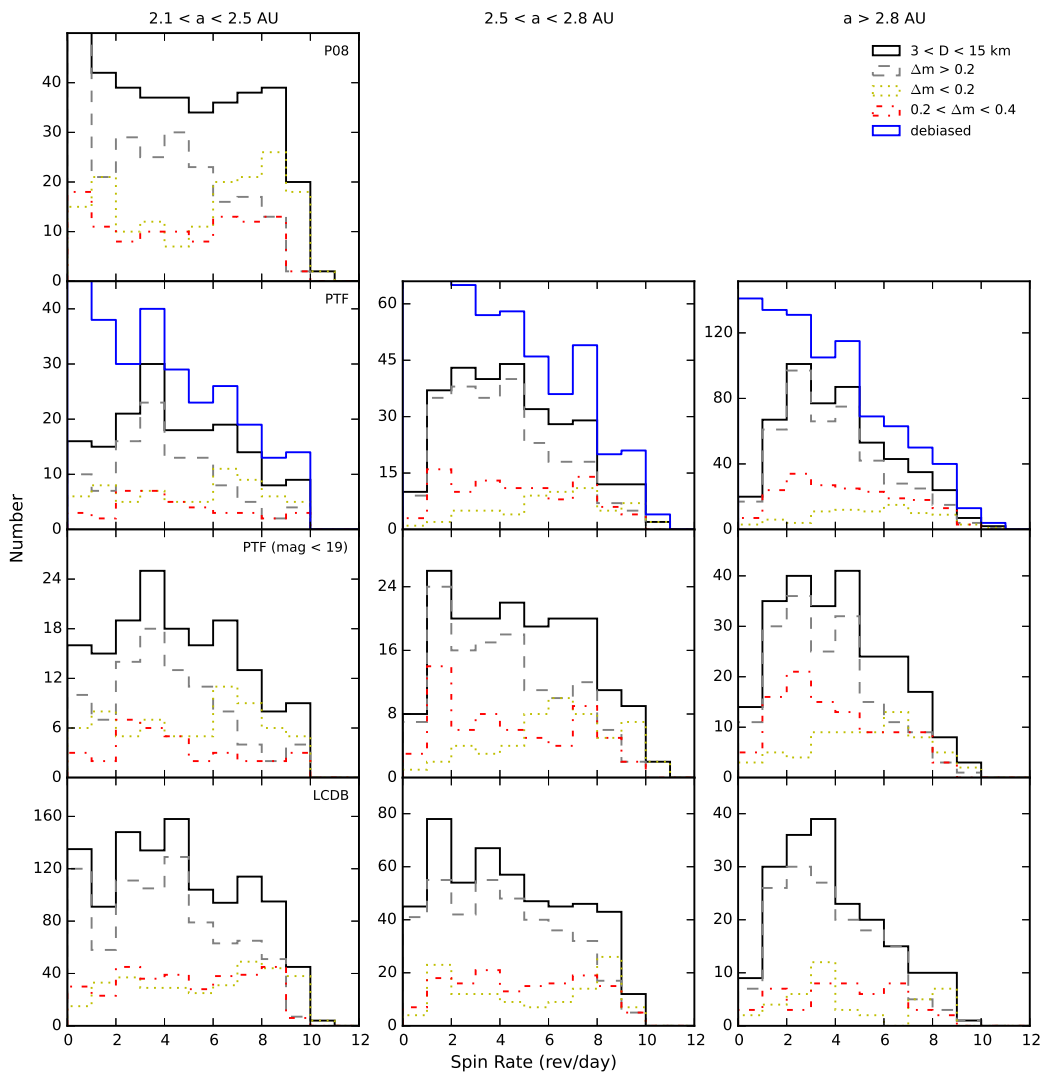}
  \caption{Spin-rate distributions of D3-15 asteroids (black line) in inner (left), mid (middle) and outer (right) main
  belts for P08 (update 2014-04-20; first row), PTF-U2s (second row), PTF-U2s with magnitude $< 19$ mag (third row)
  and LCDB (last row). The blue solid, gray dashed, yellow dotted and red dot-dashed lines represent
  debiased PTF-U2s, asteroids with $\Delta m > 0.2$, $\Delta m < 0.2$ and $0.2 < \Delta m < 0.4$ mag, respectively.}
  \label{08_3}
  \end{figure}

To have compatible diameter range with P08 (update 2014-04-20), we use asteroids of $3 < D < 15$ km
(hereafter, D3-15) to compare the spin-rate distributions in Fig.~\ref{spin_rate_comp}. In contrast to
P08 (update 2014-04-20), none of PTF-U2s, M08 and LCDB have a flat distribution, which is
quasi-Maxwellian for PTF-U2s (i.e., a peak at $3 \leq f \leq 5$ rev/day and a slow decrease after) and
looks like a step function with a number decrease at $f = 6$ rev/day for M09 and LCDB. With debiased
PTF-U2s, we cannot obtain a distribution as flat as P08 (update 2014-04-20) and the number decrease
still remains. When these samples are divided by $\Delta m = 0.2$ mag, the spin-rate distributions
become more consistent with each other for asteroids with $\Delta m > 0.2$ mag (i.e.,
quasi-Maxwellian). To have a further investigation, we separate them into inner ($2.1 < a < 2.5 AU$),
mid ($2.5 < a < 2.8$ AU) and outer ($a > 2.8$ AU) main belt for a detail investigation. Note that M09
is no longer included in the following analysis due to its insufficient number of sample for finer
parameter space. Fig.~\ref{08_3} shows the difference between P08 and the others are mainly from
asteroids with $\Delta m < 0.2$ mag where P08 (update 2014-04-20) has relative more fast rotators
(i.e., $f \ge 6$ rev/day). We believe this discrepancy could be primarily due to different survey
strategies as pointed out by \citet{Masiero2009}. However, this requests a more comprehensive study.
Although the quasi-Maxwellian distribution could possibly be an observational bias due to a greater
ability in detecting $2 \le f \le 5$~rev/day \citep{Harris2012}, we do not see this situation in the
spin-rate distributions for various intervals of $\Delta m$ (i.e., roughly flat for asteroids with
$\Delta m < 0.2$ and $0.2 < \Delta m < 0.4$ mag). Moreover, most of asteroids with $\Delta m \ge 0.4$
mag are of $2 \le f \le 5$~rev/day and only a few at $f \ge 6$ rev/day that consists with the sense of
lower spin-rate limit for large-amplitude asteroid (see Fig.~\ref{spin_amp}). Moreover, our detection
simulation shows roughly fair detection rate for $f > 2$ rev/day. This indicates that the
quasi-Maxwellian distribution for D3-15 asteroids is not a result of favoring to detect asteroids of
$2 \le f \le 5$~rev/day.

\begin{figure}
  \epsscale{1}
  \plotone{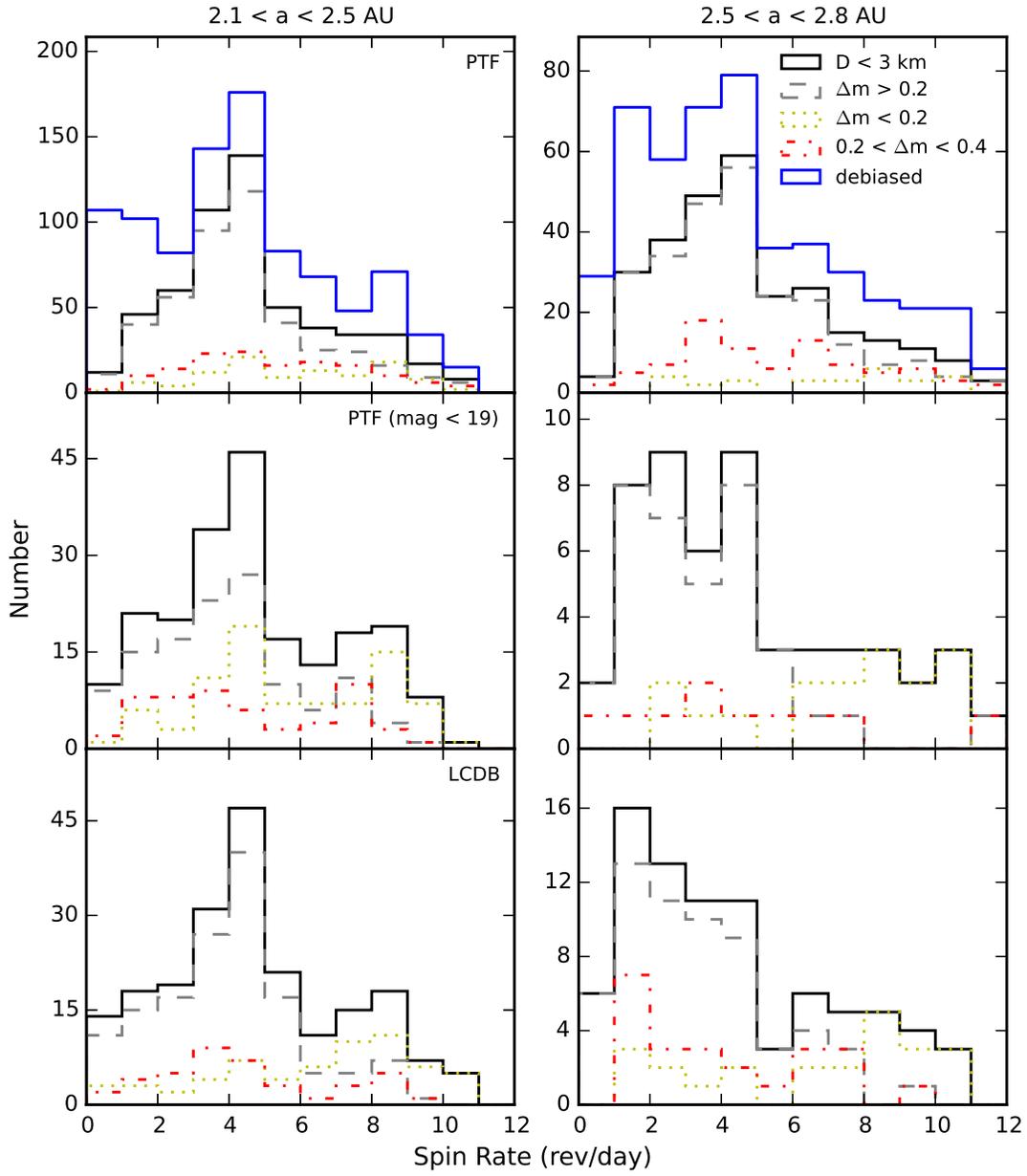}
  \caption{Spin-rate distributions of asteroids with $D < 3$ km. Same as Fig.~\ref{08_3}.}
  \label{08_5}
  \end{figure}

When we look at the spin-rate distribution of asteroids with $D < 3$ km in Fig.~\ref{08_5}, we notice
an obvious number decrease at $f = 6$ rev/day both in PTF-U2s and debiased PTF-U2s. This number drop
still remains when applying a amplitude limit of $\Delta m > 0.2$ mag. Since the YORP effect timescale
is relatively shorter for asteroids of $D < 3$ km and moreover the spin-rate limit is lower for
large-amplitude asteroids, we believe that more large-amplitude asteroids of $D < 3$ km have been spun
up to reach their spin-rate limit and broken up when comparing with D3-15 asteroids that results in a
number decrease at $f = 6$ rev/day. Note that the small numbers at $f < 3$ rev/day in PTF-U2s is due
to the lower detection rate for small-sized asteroids with long period. However, this deficiency
becomes more comparable after debiasing.

Our result does not affected much by applying a brighter limiting magnitude of 19 mag (see bottom rows
in Fig.~\ref{08_3} and Fig.~\ref{08_5}), such that our conclusion remains the same.

\subsubsection{Spin Rate vs.\ Spectral Type}
\begin{figure}
  \epsscale{1}
  \plotone{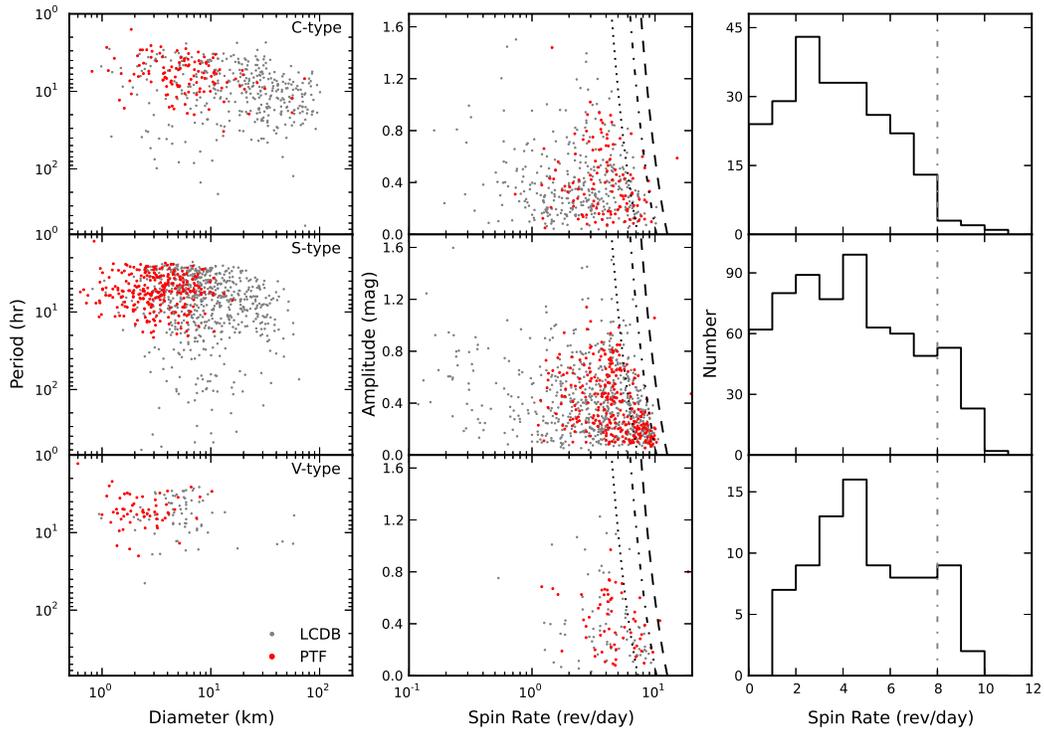}
  \caption{Asteroid rotation period vs.\ diameter (left), lightcurve amplitude vs.\ spin rate (middle),
   and spin-rate distribution (right) for C- (upper), S- (middle) and V-type (lower) asteroids. The gray and red filled
  circles are LCDB objects and PTF-U2s, respectively.} 
  \label{dia_per_tax}
  \end{figure}

In PTF-U2 and LCDB data sets, there are 478 C-type, 928 S-type, and 136 V-type asteroids according to
SDSS colors. When we look at the plot of diameter vs.\ rotation period (left panel in
Fig.~\ref{dia_per_tax}) and the plot of spin rate vs.\ lightcurve amplitude (middle panel) for the C-,
S-, and V-types (upper, middle and lower panels, respectively), all of them occupy similar regions in
rotation-period or spin-rate spaces. Because of the limiting magnitude of the observations and the
various geometric albedo values involved, the asteroids have different diameter ranges. We note that
C- and V-type asteroids seldom have objects with $P > 100$~hr and $P > 20$~hr, respectively. We see
the S-type and V-type show clear boundaries at $\rho = 2$ g/cm$^3$, and the C-type seems to have a
boundary at $\rho = 1.5$ g/cm$^3$ with a small group extending to $\rho = 2$ g/cm$^3$. Therefore, the
C-type asteroids have much fewer objects with $f > 8$~rev/day. Since M-type and E-type asteroids
cannot be separated from C-type asteroids by SDSS colors, the small group in the C-type asteroids is
properly due to those indistinguishable M-type and E-type asteroids, which have larger bulk density
with respect to C-type. Despite the small group of C-type, what we see here is in a good agreement
with the general sense of a lower bulk density for C-type relative to S-/V-type. For a comparison
without possible incompleteness at the large/small-sized ends, we controlled our samples in the range
of $3 < D < 20$~km to generate spin-rate distributions for these three asteroid types (right panel of
Fig.~\ref{dia_per_tax}), in which there were 229, 657, and 81 objects, respectively. The overall
shapes do not appear to be the same, but all distributions peak at $2 \le f \le 5$~rev/day. Besides,
the C-type distribution is more Maxwellian-like. The Kolmogorov-Smirnov test on the spin-rate
distribution gives $p-$values of 0.01, 0.003, and 0.026 for each pairing of CS, CV and SV types,
respectively. This indicates that their spin-rate distributions might be different to each other.
However, this preliminary study needs to be confirmed when the samples are more complete in the
parameter space.

\section{Summary}
Two surveys for asteroid rotation periods have been carried out by using the iPTF in January 6-9 and
February 20-23 of 2014. We obtained 1,438 rotation periods from this campaign having quality codes $U
\geq 2$, and most of them are associated with main-belt asteroids. There are 53 survey objects that
match objects in the LCDB, and the derived rotation periods between these data sets are in mostly good
agreement, an indication that our analysis is reliable.

Integrating this result with our previous study \citep{Chang2014a}, we found the spin-rate
distributions for D3-15 asteroids of PTF-U2s, M09 and LCDB are quasi-Maxwellian that shows number
decrease with frequency for $f > 5$ rev/day. The discrepancy between P08 (update 2014-04-20) and the
others is mainly from asteroids with $\Delta m < 0.2$ mag that properly due to different approaches of
acquiring samples of asteroid rotation period. In addition, we found a significant number drop at $f =
6$ rev/day for asteroids of $D < 3$ km. This might be explained by YORP effect which works faster on
small-sized asteroids and push those elongate objects over their lower spin-rate limit to create this
number drop at $f = 6$ rev/day.

Along with a confirmed large-sized SFR, 2005 UW163 \citep{Chang2014b}, we also found other five SFR
candidates. If $P \sim 3.3 \sqrt{(1+\Delta m)/\rho}$ is followed for these asteroids, then their bulk
densities are all much larger than 3~g/cm$^3$. This suggests that cohesiveness and tensile strength,
in addition to self-gravity, should also play roles in keeping them from breaking apart under such
fast rotation.

With the available SDSS colors, C-, S- and V-type asteroids might have different distributions.
However, their Maxwellian-like distributions suggests that they still retain certain degree of
collisional equilibrium. Moreover, we note that evidence suggests that C-type asteroids have a lower
spin-rate limit than the S-type. This agrees with the general sense that C-type asteroids have a lower
bulk density than the S-type.

\acknowledgments This work is supported in part by the National Science Council of Taiwan under the
grants NSC 101-2119-M-008-007-MY3 and NSC 102-2112-M-008-019-MY3. We thank the referees, Petr Pravec
and Alan Harris, for their useful suggestions and comments to improve the content of the paper. This
publication makes use of data products from $WISE$, which is a joint project of the University of
California, Los Angeles, and the Jet Propulsion Laboratory/California Institute of Technology, funded
by the National Aeronautics and Space Administration. This publication also makes use of data products
from $NEOWISE$, which is a project of the Jet Propulsion Laboratory/California Institute of
Technology, funded by the Planetary Science Division of the National Aeronautics and Space
Administration. We gratefully acknowledge the extraordinary services specific to $NEOWISE$ contributed
by the International Astronomical Union's Minor Planet Center, operated by the Harvard-Smithsonian
Center for Astrophysics, and the Central Bureau for Astronomical Telegrams, operated by Harvard
University.

\clearpage

\clearpage

\clearpage

\clearpage

\clearpage

\clearpage

\clearpage
\begin{deluxetable}{lrrcccc}
\tabletypesize{\scriptsize} \tablecaption{Survey observations in January 2014. \label{obs_log_01}}
\tablewidth{0pt}
\startdata \tableline\tableline
Field ID & RA & Dec. & Jan 6 & Jan 7 & Jan 8 & Jan 9 \\
         & ($^{\circ}$) & ($^{\circ}$) & $\Delta$t, N$_\textrm{exp}$ & $\Delta$t, N$_\textrm{exp}$ & $\Delta$t, N$_\textrm{exp}$ & $\Delta$t, N$_\textrm{exp}$ \\
\tableline
    3559 &  117.00  &  19.12 &     9.6, 28 &    9.8, 30 &    5.1, 16 &    9.7, 12\\
    3560 &  120.60  &  19.12 &     9.7, 29 &    9.8, 30 &    5.1, 16 &    9.7, 12\\
    3561 &  124.20  &  19.12 &     9.3, 28 &    9.4, 29 &    5.5, 17 &    9.6, 11\\
    3562 &  127.80  &  19.12 &     9.3, 28 &    9.1, 28 &    5.5, 17 &    8.2, 11\\
    3563 &  131.40  &  19.12 &     9.2, 28 &    8.8, 27 &    5.1, 16 &    8.2,  9\\
    3564 &  135.00  &  19.12 &     8.9, 27 &    8.7, 27 &    5.1, 16 &    8.2,  9\\
\enddata
\tablecomments{$\Delta$t is the time duration spanned by each observing set in hours and
N$_\textrm{exp}$ is the total number of exposures for each night and field.}
\end{deluxetable}

\begin{deluxetable}{llrrrrrrrrrrlrrrrl}
\tabletypesize{\scriptsize} \setlength{\tabcolsep}{0.02in} \tablecaption{Synodic rotation periods of
PTF-U2s. \label{table_p}} \tablewidth{0pt} \tablehead{ \colhead{Obj ID} & \colhead{Designation} &
\colhead{$a$} & \colhead{$e$} & \colhead{$i$} & \colhead{$\Omega$} & \colhead{$\omega$} & \colhead{D}
& \colhead{$\triangle$} & \colhead{$r$} & \colhead{$\alpha$} & \colhead{$H_R$} & \colhead{n} &
\colhead{m} & \colhead{$PTF_R$} & \colhead{Period (hr)} & \colhead{$\triangle m$} & \colhead{U}}
\startdata
00414\tablenotemark{*}         & (414) Liriope                &  3.52 &  0.07 &  9.56 & 110.6 & 322.7 &  88.8\tablenotemark{w}   &   3.32 &   2.35 &  3.38 &  9.22$\pm$0.12   &  4 &  81 & 13.99$\pm$0.00   &   7.38$\pm$0.11  &  0.14 &  2 \\
00659\tablenotemark{*}         & (659) Nestor                 &  5.19 &  0.11 &  4.52 & 350.9 & 342.0 & 122.3\tablenotemark{w}   &   5.78 &   4.79 &  0.70 &  8.47$\pm$0.13   &  4 &  89 & 15.78$\pm$0.00   &  16.00$\pm$0.26  &  0.22 &  3 \\
01098\tablenotemark{*}         & (1098) Hakone                &  2.69 &  0.12 & 13.38 & 329.0 &  81.4 &  29.6\tablenotemark{w}   &   2.70 &   1.72 &  2.89 & 10.35$\pm$0.12   &  2 &  43 & 13.94$\pm$0.00   &   7.16$\pm$0.05  &  0.35 &  3 \\
01117\tablenotemark{*}         & (1117) Reginita              &  2.25 &  0.20 &  4.34 & 147.1 & 151.1 &  10.2\tablenotemark{w}   &   2.58 &   1.60 &  1.03 & 11.47$\pm$0.07   &  2 &  48 & 14.70$\pm$0.00   &   2.94$\pm$0.01  &  0.11 &  3 \\
01153\tablenotemark{*}         & (1153) Wallenbergia          &  2.20 &  0.16 &  3.34 & 280.6 &  28.9 &   9.9       &   2.52 &   1.54 &  3.64 & 12.04$\pm$0.10   &  1 &  22 & 15.22$\pm$0.00   &   4.12$\pm$0.07  &  0.23 &  2 \\
01254\tablenotemark{*}         & (1254) Erfordia              &  3.14 &  0.03 &  7.07 & 288.0 & 239.4 &  56.1\tablenotemark{w}   &   3.08 &   2.10 &  1.88 & 10.23$\pm$0.12   &  3 &  68 & 14.58$\pm$0.00   &  12.31$\pm$0.16  &  0.33 &  3 \\
01443\tablenotemark{*}         & (1443) Ruppina               &  2.94 &  0.06 &  1.93 & 174.9 & 163.7 &  17.2\tablenotemark{w}   &   3.12 &   2.14 &  2.33 & 10.97$\pm$0.12   &  4 &  85 & 15.21$\pm$0.00   &   5.89$\pm$0.04  &  0.27 &  3 \\
01470\tablenotemark{*}         & (1470) Carla                 &  3.16 &  0.07 &  3.21 & 358.5 & 342.5 &  34.1\tablenotemark{w}   &   3.37 &   2.38 &  0.89 & 10.80$\pm$0.12   &  4 & 107 & 15.49$\pm$0.00   &   6.15$\pm$0.04  &  0.24 &  3 \\
\enddata
\tablecomments{Columns: asteroid's designations, semi-major axis ($a$, AU), eccentricity ($e$,
degree), inclination ($i$, degree), longitude of ascending node ($\Omega$, degree), argument of
periapsis ($\omega$, degree), diameter (D, km), heliocentric distance ($\triangle$, AU), geodesic
distance ($r$, AU), phase angle ($\alpha$, degree), absolute magnitude ($H$, mag), number of nights
(n), number of images (m), $PTF_R$ magnitude, derived rotation period (hours), lightcurve amplitude
(mag) and rotation period quality code (U).  The full table is available in the eletronic version.}
\tablenotetext{*}{Asteroid available in the LCDB.} \tablenotetext{b}{Lightcurves with large amplitudes
and deep V-shape minima.} \tablenotetext{w}{$WISE$/$NEOWISE$ diameter.} \tablenotetext{e}{Estimated
diameter from Eq~\ref{dia_eq}.}
\end{deluxetable}

\begin{deluxetable}{llrrrrrrrrrrlrrrrl}
\tabletypesize{\scriptsize} \setlength{\tabcolsep}{0.02in} \tablecaption{Asteroids with partial phase
coverage. \label{table_p_part}} \tablewidth{0pt} \tablehead{ \colhead{Obj ID} & \colhead{Designation}
& \colhead{$a$} & \colhead{$e$} & \colhead{$i$} & \colhead{$\Omega$} & \colhead{$\omega$} &
\colhead{D} & \colhead{$\triangle$} & \colhead{$r$} & \colhead{$\alpha$} & \colhead{$H_R$} &
\colhead{n} & \colhead{m} & \colhead{$PTF_R$} & \colhead{Period (hr)} & \colhead{$\triangle m$} &
\colhead{U}} \startdata
00850\tablenotemark{*}         & (850) Altona                 &  3.01 &  0.12 & 15.53 & 121.2 & 133.5 &  59.9\tablenotemark{w}   &   3.24 &   2.30 &  6.02 &  9.37$\pm$0.13   &  4 &  86 & 14.11$\pm$0.00   &  11.16$\pm$0.38  &  0.10 &  2 \\
01100\tablenotemark{*}         & (1100) Arnica                &  2.90 &  0.07 &  1.03 & 304.2 &  23.8 &  17.2\tablenotemark{w}   &   3.06 &   2.08 &  3.05 & 10.39$\pm$0.11   &  4 &  79 & 14.77$\pm$0.00   &  14.55$\pm$0.22  &  0.09 &  2 \\
01135\tablenotemark{*}         & (1135) Colchis               &  2.67 &  0.12 &  4.54 & 350.8 &   4.1 &  49.8\tablenotemark{w}   &   2.95 &   1.96 &  1.28 & 10.26$\pm$0.18   &  4 &  77 & 14.21$\pm$0.00   &  23.41$\pm$1.09  &  0.33 &  2 \\
01449\tablenotemark{*}         & (1449) Virtanen              &  2.22 &  0.14 &  6.64 & 110.8 & 131.9 &   9.9\tablenotemark{w}   &   2.36 &   1.42 &  9.32 & 11.69$\pm$0.15   &  4 &  64 & 14.92$\pm$0.00   &  14.77$\pm$0.44  &  0.27 &  2 \\
02009\tablenotemark{*}         & (2009) Voloshina             &  3.12 &  0.14 &  2.86 & 107.6 &   6.3 &  26.6\tablenotemark{w}   &   2.75 &   1.76 &  1.27 & 10.87$\pm$0.12   &  4 &  88 & 14.59$\pm$0.00   &   2.94$\pm$0.01  &  0.32 &  2 \\
03014\tablenotemark{*}         & (3014) Huangsushu            &  2.36 &  0.23 &  0.99 & 140.8 & 180.6 &   6.9\tablenotemark{w}   &   2.89 &   1.90 &  0.78 & 13.15$\pm$0.15   &  3 &  68 & 16.96$\pm$0.01   &  24.00$\pm$0.62  &  0.35 &  2 \\
03031\tablenotemark{*}         & (3031) Houston               &  2.24 &  0.10 &  4.34 & 317.8 & 249.1 &   6.8\tablenotemark{w}   &   2.11 &   1.12 &  4.17 & 12.81$\pm$0.08   &  4 &  62 & 15.06$\pm$0.00   &   5.61$\pm$0.06  &  0.17 &  2 \\
03033\tablenotemark{*}         & (3033) Holbaek               &  2.24 &  0.09 &  4.74 & 166.9 &  67.3 &   9.1\tablenotemark{w}   &   2.19 &   1.20 &  1.81 & 12.68$\pm$0.09   &  4 &  82 & 14.97$\pm$0.00   &  20.00$\pm$0.87  &  0.15 &  2 \\
\enddata
\tablecomments{The amplitudes of the objects with partial lightcurve coverage and lightcurves with a
single minimum should be treated as lower limits.  Also, see note and footnotes associated with
Table~\ref{table_p} for nomenclature and explanation.}
\end{deluxetable}

\begin{deluxetable}{llrrrrrrrrrrlrrrrl}
\tabletypesize{\scriptsize} \setlength{\tabcolsep}{0.02in} \tablecaption{Bright asteroids with large
lightcurve variation  and without rotation-period determination. \label{table_nd}} \tablewidth{0pt}
\tablehead{ \colhead{Obj ID} & \colhead{Designation} & \colhead{$a$} & \colhead{$e$} & \colhead{$i$} &
\colhead{$\Omega$} & \colhead{$\omega$} & \colhead{D} & \colhead{$\triangle$} & \colhead{$r$} &
\colhead{$\alpha$} & \colhead{$H_R$} & \colhead{n} & \colhead{m} & \colhead{$PTF_R$} &
\colhead{$\triangle m$}} \startdata
02979             & (2979) Murmansk              &  3.12 &  0.16 & 11.42 & 148.5 & 197.2  &  18.5\tablenotemark{w}   &   3.60 &   2.62 &  1.55 & 12.35$\pm$0.18   &  4 &  83 & 17.41$\pm$0.02   &  0.60 \\
03427             & (3427) Szentmartoni          &  2.28 &  0.13 &  2.60 & 310.1 &  76.4  &   5.0\tablenotemark{w}   &   2.43 &   1.44 &  1.01 & 13.69$\pm$0.24   &  4 &  84 & 16.55$\pm$0.01   &  0.75 \\
04251             & (4251) Kavasch               &  2.41 &  0.18 &  3.34 & 115.4 & 132.6  &   4.3\tablenotemark{w}   &   2.67 &   1.69 &  2.77 & 13.62$\pm$0.17   &  4 &  69 & 17.20$\pm$0.01   &  0.66 \\
04869             & (4869) Piotrovsky            &  2.24 &  0.17 &  3.55 &  96.3 & 217.6  &   6.5\tablenotemark{w}   &   2.60 &   1.64 &  6.01 & 12.93$\pm$0.32   &  4 &  78 & 16.57$\pm$0.01   &  0.83 \\
05827             & (5827) Letunov               &  2.19 &  0.13 &  3.70 & 105.4 & 284.0  &   5.0       &   2.16 &   1.25 & 12.32 & 13.53$\pm$0.32   &  4 &  74 & 16.38$\pm$0.01   &  0.97 \\
07321             & (7321) 1979 MZ2              &  2.53 &  0.09 &  2.29 & 154.5 & 328.0  &   5.2\tablenotemark{w}   &   2.34 &   1.35 &  3.20 & 13.72$\pm$0.22   &  4 &  83 & 16.44$\pm$0.01   &  0.80 \\
08475             & (8475) Vsevoivanov           &  3.08 &  0.25 &  4.83 & 118.5 & 266.1  &  14.3\tablenotemark{w}   &   2.85 &   1.87 &  2.59 & 12.92$\pm$0.25   &  4 &  79 & 16.75$\pm$0.01   &  0.64 \\
09319             & (9319) 1988 RV11             &  2.25 &  0.15 &  2.94 & 165.5 &  97.5  &   5.5\tablenotemark{w}   &   2.32 &   1.33 &  1.02 & 14.87$\pm$0.18   &  4 &  85 & 17.53$\pm$0.02   &  0.66 \\
\enddata
\tablecomments{$\Delta m$ is the lightcurve variation over our four-night observing-time span. Also,
see note and footnotes associated with Table~\ref{table_p} for nomenclature and explanation.}
\end{deluxetable}

\end{document}